\documentclass[aps,twocolumn,preprintnumbers,nofootinbib,floatfix,superscriptaddress]{revtex4}
\usepackage{eurosym}
\usepackage{amsmath}
\usepackage{graphicx}
\usepackage{amsfonts}
\usepackage{array}
\usepackage{amsthm}
\usepackage{bm}
\usepackage{palatino}
\usepackage{mathpazo}
\usepackage{supertabular}
\usepackage{subfig}
\usepackage[breaklinks]{hyperref}
\usepackage[font=small,labelfont=bf,
justification=justified,
format=plain]{caption}
\usepackage[font=small,labelfont=bf,justification=justified]{caption}
\usepackage{color}
\usepackage{booktabs}
\usepackage{multirow}
\usepackage[labelfont=bf,labelsep=quad,justification=justified]{caption}

\newcommand{\be}{\begin{equation}}
\newcommand{\ee}{\end{equation}}
\newcommand{\ba}{\begin{eqnarray}}
\newcommand{\ea}{\end{eqnarray}}
\newcommand{\bal}{\begin{align}}
\newcommand{\eal}{\end{align}}

\newcommand{\bw}{\begin{widetext}}
\newcommand{\ew}{\end{widetext}}

\begin{document}
\title{Accretion of matter onto black holes in massive gravity with Lorentz symmetry breaking}
\author {Abdul Jawad}
\email{abduljawad@cuilahore.edu.pk}
\affiliation{ Department of Mathematics, COMSATS University
Islamabad, Lahore-Campus, Lahore-54000, Pakistan.}
\author{Kimet Jusufi}
\email{kimet.jusufi@unite.edu.mk}
\affiliation{Physics Department, State University of Tetovo, Ilinden Street nn, 1200,
Tetovo, North Macedonia}
\author{M. Umair Shahzad}
\email{mushahzad@uo.edu.pk}
\affiliation{ Department of Mathematics, University of Okara, Okara, Pakistan.}

\begin{abstract}
In this paper we study the accretion of matter onto the black holes and the 
shadow images obtained by an infalling accretion flow of gas in a theory of massive gravity (MG) with a spontaneously breaking Lorentz
symmetry. This black hole solution is characterized by mass $M$, scalar charge $S$ and the parameter $\lambda$. In order to extract the astrophysical results of our analyses, firstly, we have assumed a specific range for the parameter $\lambda$ to constrain the scalar charge $S$ using the EHT result. To this end, we have studied the effect of the scalar charge on the intensity of the electromagnetic radiation from the black hole. Moreover, we investigate the behavior of polytropic as well as the isothermal fluid flow onto massive gravity black hole and notice that accretion starts from supersonic/subsonic flow, passes through the critical point using particular model parameter and ends near the horizon. We also analyzed the mass accretion rate in the presence of various fluids which indicates important signatures.  We also elaborate on the possibility to analyze the phase transition and the stability of the black hole using the shadow formalism.
\end{abstract}

\pacs{}
%\keywords{...}
\maketitle
\section{Introduction} 
During the last decades a lot of efforts has been made to modify general relativity and, in particular, to formulate a theory with massive gravitons known as massive gravity theory. Such a theory is speculated to have rich phenomenology, for example it may explain many problems is cosmology without invoking the concept of dark energy.  Initially it was formulated a linear theory of MG by Fierz and Pauli \cite{fp}, but unfortunately it was shown that this theory suffered from the so-called discontinuity which was subsequently addressed by Vainshtein \citep{v}. Furthermore it was also shown to exists another issue in MG, namely a ghost instability \cite{bd}, which then lead to further extension of this theory \cite{de}. Note that recently other formulations of MG have been proposed \cite{oth}. It is interesting that a relation between the hierarchy problem and brane-world theories with the MG was investigated in Ref. \cite{bw}. Recently, Bebronne and Tinyakov \cite{bebronne}, found an interesting black hole solution in MG which depends on additional parameter known as the scalar charge $S$. In this theory the deflection of light  was studied in Ref. \cite{Jusufi:2017drg}, strong lensing \cite{sl}, shadows in rotating regular BH in MG was investigated in Ref. \cite{Jusufi:2019caq}, phase transition  and thermodynamics \cite{ft,td}, quasi-normal modes \cite{qnm} and references therein.  

Nowadays it is widely known that the process of accretion of matter onto a black hole plays significant role in explaining many astrophysical phenomena. In particular, the accretion of gas onto the black holes is linked  to the quasi-periodic oscillations observed by astronomers. On the other hand, the emission of electromagnetic radiation in the form of gamma-ray burst, X-ray binaries and the tidal disruption events are manly related to the process of accretion matter onto black holes. 

In this work, we would like to understand to what extant the effect of massive gravity with a spontaneously breaking Lorentz symmetry are important in the accretion of matter onto black holes and investigate the shadow images. In order to have a more realistic picture, we are going to use the EHT result for the  M87 black hole to constrain the scalar charge parameter in this theory. We aim to construct the shadow images and the intensity of radiating infalling gas (see related works with different accretion models, \cite{Narayan}-\cite{Zeng}). To this end, we are going to investigate the matter accretion, we are going to use a polytropic fluid flow and address the critical points of the accretion process obtained by different fluids such as: sub-relativistic fluid, radiation fluid, ultra-relativistic fluid and ultra-stiff fluid. In this regard, beside the effect of black hole mass it is interesting to investigate the effect of "hair parameter" ($\lambda$) and scalar charge $S$ on the change of mass or accretion rate using a particular domain of parameters.  For studies related to the accretion of matter onto black holes see the following Refs. (\cite{19}-\cite{Farooq:2020lgm}).

The rest of this paper is structured. In Sec. \textbf{II} we shortly review the basics of the BH solution in MG. In  Sec. \textbf{III} we study the BH shadow using a radiation thin accretion flow and the intensity of infalling radiating gas.  In  Sec. \textbf{IV} we use the EHT  result to constrain the scalar charge $S$ and the parameter $\lambda$. In  Secs. \textbf{V} and \textbf{VI}, we investigate the accretion rate using fluid with different EoS such as the polytropic fluid using different fluids. In Sec. \textbf{VII}  we elaborate on the connection between the phase transition and the shadow radius.  Finally in Sec. \textbf{VIII} we comment on our results. In the present paper we set the natural units $G=c=\hbar=1$.

\section{Black holes in massive gravity}
In the present work we shall consider the massive gravity model described by the following action \cite{bebronne}
\begin{equation} \mathcal{S}_{MG} = \int  d^4x  \sqrt{-g} \left[ \frac{R}{16 \pi }+ \Lambda^4 \mathcal{F}(X, W^{ij}) \right],
\end{equation}
with $R$ being the scalar curvature and function $\mathcal{F}$ which describes the scalar fields $\psi^i$ and $\psi^0$,  respectively.  Note that the scalar field is minimally coupled to gravity while in the theory the Lorentz symmetry is spontaneously broken.   The parameter $\Lambda$ is proportional to $\sqrt{m}$, with $m$ being the graviton
mass.
It is possible to express the function $\mathcal{F}$ in terms of two Goldstone fields, $X$ and $W^{ij}$, given by
\begin{equation} 
X = \frac{\partial^0 \psi^i\partial_0\psi^i}{\Lambda^4},
\end{equation}
\begin{equation}
W^{ij} = \frac{\partial^\mu \psi^i\partial_\mu \psi^j}{\Lambda^4}-\frac{\partial^\mu \psi^i\partial_\mu \psi^0 \partial^\nu \psi^j\partial_\nu \psi^0}{\Lambda^4 X}.
\end{equation}
It is important to note that finding analytical solutions for a generic function $\mathcal{F}$ is not possible. One can choose this function in such a way that the
resulting equations are solvable analytically. Using a static and spherically symmetric metric (detailed derivation can be found in \cite{bebronne})
 \begin{equation} \label{metric1}
ds^2=- f(r) dt^2+g(r)dr^2+r^2\left(d\theta^2+\sin^2\theta d\phi^2   \right),
\end{equation}
with 
\begin{equation} 
\psi^0 = \Lambda^2(t+N(r) ),\,\,\,\, \psi^i= \phi(r) \frac{\Lambda^2x^i}{r},
\end{equation}
along with the choice \cite{bebronne}
\begin{equation}
\mathcal{F}=c_0\left(\frac{1}{X}+w_1\right)+c_1(w_1^3-2w_1 w_2-6w_1+2w_3-12),
\end{equation}
where $c_0$ and $c_1$ are some dimensionless constants and 
\begin{equation}
w_1=-(f_1+2f_2),\,\,\,w_2=f_1^2+2f_2^2,\,\,w_3=-(f_1^3+2f_2^3),
\end{equation}
with
\begin{equation}
f_1=\frac{{\phi'}^2}{f g X},\,\,\,f_2=\frac{\phi^2}{r^2},\,\,\,X=\frac{g-f {N'}^2 }{f g}.
\end{equation}
Solving the field equations the solution is found to be \cite{bebronne}
  \begin{equation} 
f(r)= 1-\frac{r_s}{r}-\frac{S}{r^{\lambda}}+\Lambda_c r^2, 
\end{equation}
in which $r_s$ and $S$ are integrating constants, and
\begin{equation}
    \lambda=-12 b^6 \frac{c_1}{c_0} ,\,\,\Lambda_c=2m^4 c_1(b^6-1).
\end{equation}
Furthermore it was shown that the following equations holds \cite{bebronne}
\begin{eqnarray}\nonumber
g(r)&=&\frac{1}{f(r)},\\\nonumber
N(r) &=& \pm \int\frac{dr}{f(r)} \left[ 1-f(r) \left( \frac{S\lambda (\lambda-1)}{c_0 m^2} \frac{1}{r^{\lambda+2}} +1\right)^{-1}\right]^{\frac{1}{2}},\\
\phi(r)&=& b r.
\end{eqnarray}
Note that $b$ is related to $c_0$ and $c_1$, and satisfies the following equation \cite{bebronne}
\begin{eqnarray}
(b^2-1)(6b^4+6 b^2+c_0/c_1)=0.
\end{eqnarray}
 We can see that if $b=1$ and $c_1/c_0>0$, then $\Lambda_c=0$ and $\lambda<0$. Then the black hole metric is not asymptotically flat, if $c_1/c_0<0$ then $\lambda>0$. However one can further observe that when $\lambda < 1$ the the ADM mass of these solutions is infinite. Therefore to recover a spacetime with a finite ADM mass we shall focus on the case $\lambda\geq 1$ and set $b=1$. In such a case the solution reduces to 
\begin{equation} 
f(r)= 1-\frac{2M}{r}-\frac{S}{r^{\lambda}}, 
\end{equation}
where in the limit of vanishing scalar charge $S$ the Schwarzschild black hole solution is obtained. We have identified $r_s=2M$, and $M$ is the AMD mass provided $\lambda>1$. We can see this fact by using the following definition for the ADM mass \cite{Shaikh:2018kfv}
\begin{equation}
    M=\lim_{r\to \infty} \frac{r}{2}\left[\frac{1-f(r)}{f(r)}\right]\simeq M+\frac{S}{2  r^{\lambda-1}}.
\end{equation}
If we set $\lambda=1$, the ADM  mass shifts by a constant and we can say that such a black hole cannot be distinguished from the Schwarzschild black hole, due to the simple scaling of mass $M \to M +S/2$. But for any $\lambda>1$ the ADM mass is exactly $M$.
The scalar charge in general can be positive, i.e., $S > 0$, or negative i.e., $S<0$. The parameter $\lambda$ is known also as the ``hair parameter''. As a special case $\lambda=2$,  and $S=-\mathcal{Q}^2$, we obtain the RN black hole spacetime, but obviously the case $\lambda=2$ for $S>0$ differs from the RN black hole.

%\begin{figure*}[t!]
%    \includegraphics[scale=0.53]{mgintensity.pdf}
%    \includegraphics[scale=0.53]{mgintensity2.pdf}
%    \caption{Left panel: The corresponding intensities using $S=0.5$ ($\lambda=1$) and $S=0.1$ ($\lambda=2$). The corresponding intensities  using $S=-0.1$ ($\lambda=1$). and $S=-0.45$ ($\lambda=2$). }
%\end{figure*}
\begin{figure*}
\centering
 \includegraphics[scale=0.73]{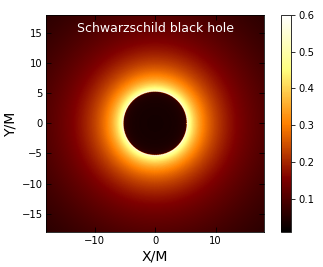}
        \includegraphics[scale=0.73]{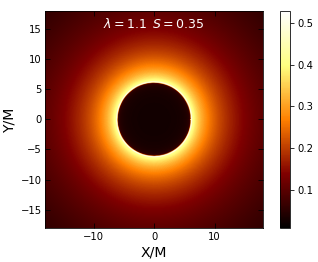}
         \includegraphics[scale=0.73]{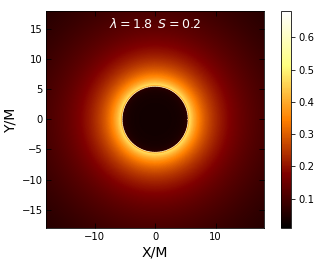}
          \includegraphics[scale=0.73]{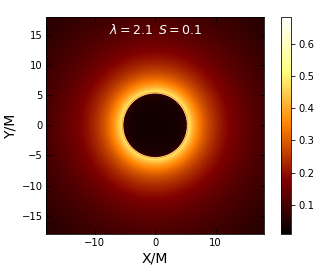}
           \includegraphics[scale=0.73]{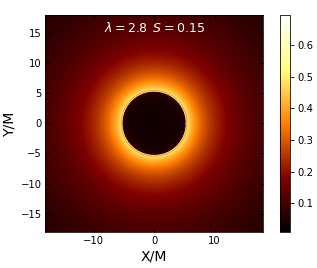}
            \includegraphics[scale=0.73]{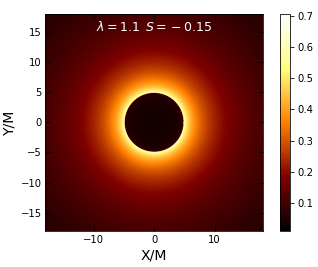}
    \caption{Images of shadows with the corresponding intensities using the infalling gas model seen by a distant observer using different values of $\lambda$ and $S$. The first plot with corresponds to the Schwarzschild black hole with $S=0$. }
\end{figure*}
\section{Black hole shadow via radiating and infalling accretion gas}
In this section we turn our attention to the shadow images and, for that purpose, let us write the Hamilton-Jacobi equation
\begin{equation}
\frac{\partial \mathcal{S}}{\partial \sigma}+\mathcal{H}=0,
\end{equation}
in the last equation $\sigma $ represents an affine parameter.  For the Hamiltonian
for light rays  it can be shown \cite{Perlick:2015vta}
\begin{equation}
\frac{1}{2}\left[-\frac{p_{t}^{2}}{f(r)}+f(r)p_{r}^{2}+\frac{p_{\phi}^{2}}{r^{2}}\right] =0.
\label{EqNHa}
\end{equation}

From the Hamiltonian one can obtain the two conserved quantities defined as
\begin{eqnarray}
p_{t}&\equiv\frac{\partial \mathcal{H}}{\partial\dot{t}}=-\mathcal{E},\\
p_{\phi}&\equiv\frac{\partial \mathcal{H}}{\partial\dot{\phi}}=\mathcal{L},
\end{eqnarray}
that is the energy $\mathcal{E}$ and angular momentum of the photon $\mathcal{L}$. The unstable orbits are obtained via the effective potential under the conditions
\begin{equation}
\mathcal{V}_{\rm eff}(r) \big \vert_{r=r_{ph}}=0,  \qquad \frac{\partial \mathcal{V}_{\rm eff}(r)}{\partial r}%
\Big\vert_{r=r_{ph}}=0,  
\end{equation}
 
One can now easily show that \cite{Perlick:2015vta}
\begin{equation}
\frac{dr}{d\phi}=\pm r\sqrt{f(r)\left[\frac{r^{2}f(r_{ph})}{r_{ph}^{2}f(r)} -1\right] }. 
\end{equation}

Let us consider that a light ray is sent from the static observer located far away from the black hole at a position $r_{0} $ along with some angle $\vartheta$ w.r.t. the radial direction. In that case, one can write \cite{Perlick:2015vta} 
\begin{equation}
\cot \vartheta =\frac{\sqrt{g_{rr}}}{g_{\phi\phi}}\frac{dr}{d\phi}\Big\vert%
_{r=r_{0}}.  \label{Eqangle}
\end{equation}

Using all these equations one can relate now the shadow radius seen by the observer and his located using the equation 
\begin{equation}
r_{s}=r_{ph}\left(\frac{1-\frac{2M}{r_0}-\frac{S}{r_0^{\lambda}}}{1-\frac{2M}{r_{ph}}-\frac{S}{r_{ph}^{\lambda}}}\right)^{1/2}\simeq \frac{r_{ph}}{\sqrt{1-\frac{2M}{r_{ph}}-\frac{S}{r_{ph}^{\lambda}}}},
\end{equation}
in which $r_{ph}$ determines the photon radius. Furthermore one can obtain the circular shape of the shadow, which can be though as a stereographic projection on the observers plane by using the coordinates, say $(X,Y)$, that are defined by
\begin{eqnarray}
X &=& \lim_{r_{0}\longrightarrow \infty} \left( -r_{0}^{2}\sin
\theta_{0}\frac{d\phi}{dr}\Big\vert_{(r_{0},\theta_{0})}\right), \\
Y &=& \lim_{r_{0}\longrightarrow \infty} \left( r_{0}^{2}\frac{d\theta}{%
dr}\Big\vert_{(r_{0},\theta_{0})}\right).
\end{eqnarray}
As we already noted that, the position of the observer far away from the black hole is given by the coordinates $( r_{0},\theta_{0}) $. In our setup, we consider here a very simple model which consists of a spherically symmetrical accretion model of infalling and radiating gas. One can now define the specific intensity $I_{\nu 0}$ which is observed at $(X,Y)$ by the observer by using the following equation  \cite{Narayan,Falce,Bambi,Shaikh,k1,k2,k3,Zeng}
\begin{equation}
    I_{obs}(\nu_{obs},X,Y) = \int_{\gamma}\mathrm{g}^3 j(\nu_{e})dl_\text{prop}.
\end{equation}

\begin{figure*}
    \includegraphics[scale=0.73]{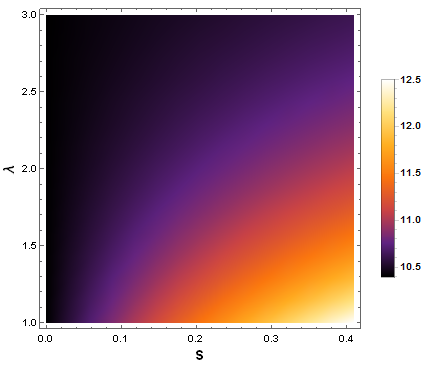}
        \includegraphics[scale=0.73]{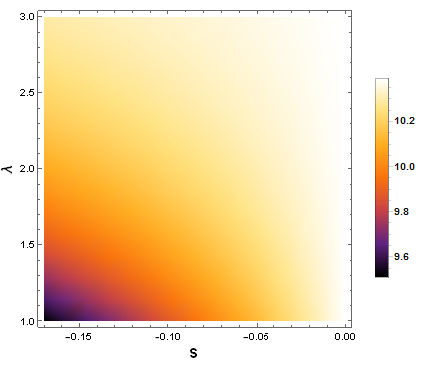}
    \caption{The regions of parameter space $(S,\lambda)$ and the diameter of the black hole. In the left panel $S>0$ in the right panel $S<0$ and both plots covers the range diameter range pf the shadow given by $9.5 \leq d_{M87} \leq 12.5$}
\end{figure*}

In the last equation we have the following important quantities: $\mathrm{g} = \nu_{obs}/\nu_{e}$ being the red-shift factor, $\nu_{e}$ ($\nu_{obs}$) is the  photon frequency which is measured in the rest-frame of the emitter (or observer), respectively. Importantly, the quantity $dl_\text{prop} = k_{\alpha}u^{\alpha}_{e}d\kappa$ measures the infinitesimal proper length, and $\kappa$ is some affine parameter. On the other hand, the quantity $j(\nu_{e})$ represents the emissivity say per unit volume in the rest-frame of the emitter and, in the present work, we are going to assume a gas such that the emission is monochromatic with emitter's-rest frame frequency given as
\begin{equation}
    j(\nu_{e}) \propto \frac{\delta(\nu_{e}-\nu_{\star})}{r^2}.
\end{equation}  

Note here that in general one can chose a different power low for the radial profile given by
\begin{eqnarray}
    j(\nu_{e}) \propto \frac{\delta(\nu_{e}-\nu_{\star})}{r^{\delta}},
\end{eqnarray}
thus, in in our case we identify $\delta=2$, but also $\delta=3$ is a commonly used case.
To obtain the red-shift factor we can use the following relation \cite{Narayan,Falce,Bambi,Shaikh,k1,k2,k3}
\begin{equation}
    \mathrm{g} = \frac{k_{\alpha}u^{\alpha}_{\text{obs}}}{k_{\beta}u^{\beta}_{e}}  ,
\end{equation}
with $k^{\mu}$ being the 4-momentum of the photon, $u^{\mu}_{\text{obs}}$ the 4-velocity of the distant observer and $u^{\alpha}_{e}$ the 4-velocity of the infalling and radiating gas which has the components (see for more details \cite{Bambi})
\begin{eqnarray}
u_e^t=\frac{1}{f(r)};\, u_e^r=-\sqrt{1-f(r)}; \,u_e^{\theta}=u_e^{\phi}=0.
\end{eqnarray}

In order to obtain the black hole images we need to integrate the intensity over all the observed frequencies, in other words we need the observed flux given by
\begin{equation}
    F_{obs}(X,Y) \propto -\int_{l_{\gamma}} \frac{\mathrm{g}^3 k_t}{r^2k^r}dr.
\end{equation}

The corresponding intensities obtained from accretion gas along with the shadow images with accretion gas by varying the hair parameter and the scalar charge are depicted in Fig. 1.  Note that we have used a region of parameters within $1\sigma$ and found that the size of the shadow radius is bigger when $S>0$ compared to the case when $S<0$. Contrary to that, we observe that the range of the intensity of the radiation is stronger for the case $S<0$.  This feature can be explained from the fact that when the scalar charge is positive, i.e., $S>0$, the deflection angle of light increases meaning that a larger number of light rays are captured by the black hole as a result the number of light rays that escape from the black hole can decrease. 

\section{Observational constraints on the scalar charge}
 In this section, we are going to constrain the scalar charge $S$ using the EHT result $\theta_s = (42 \pm 3)\mu as$, where the distance of the black hole M87 is  $D = 16.8 $ Mpc,and the mass of the M87 black hole is chosen $M = (6.5\pm 0.9) \times 10^9$ M\textsubscript{\(\odot\)}. It follows that the diameter of the shadow in units of mass $d_{M87}$ reads \cite{con}
\begin{eqnarray}
d_{M87}=\frac{D \,\theta_s}{M_{87}}=11.0 \pm 1.5.
\end{eqnarray}
Within $1\sigma $ confidence, we have the range $9.5 \leq d_{M87} \leq 12.5$, yielding the lower and upper bound of the scalar charge as follows $-0.17 \lesssim  S 	\lesssim 0.41$. This parameter space is shown in Fig. 2 where the range for $\lambda$ in our analyses is chosen to be $1 \leq \lambda \leq 3$. The diameter of the shadow is smaller for negative $S$ when the value of $\lambda$ approaches one, on the other hand, the diameter increases when $S>0$ when the value of $\lambda$ approaches one. We can see this fact from Fig. 1, for instance if we compare to the Schwarzschild black hole (case $S=0$) shadow for the values $\lambda=1.1$ and $S>0$ the shadow is considerably bigger but as the value of $\lambda$ increases the shadow radius decreases. From that figure we also the case with the smaller shadow radius obtained for $\lambda=1.1$ and $S<0$. 

\section{Polytropic Test fluid}
The motion of fluid near the BH is very important to see the particular signature of BH and there are different models which described the behavior of fluid. In this study, our particular interesting model is Chaplygin gas which leads to very
interesting results. In astrophysics, the most
general exotic fluid is modified Chalpygin gas
\cite{16}. Its equation of state is given by
\begin{equation}\label{P1}
p  = A n-\frac{B}{n^\sigma} ,
\end{equation}
where $A$ and $B$ are arbitrary constants and $0<\sigma<1$. If we
plug $A=0$, $B=-Z$ and $\sigma=-\gamma$, we get polytropic equation of state
\cite{3} i.e.
\begin{equation}
p  \equiv Z n^\gamma,
\end{equation}
where $Z$, $\gamma$ are arbitrary constants and $\gamma>0$ for
ordinary matter. Now using the technique given in Refs. (\cite{16}-\cite{34}), one can find the final form of the Hamiltonian, that is
\begin{equation}\label{P7}
H=\frac{f(r)}{1-v^2} \left(1+L\left(\frac{1-v^2}{r^4 f(r)
v^2}\right)^{\frac{\gamma-1}{2}}\right)^2,
\end{equation}
where
\begin{equation}\label{p6}
L\equiv\frac{Z \gamma n_c^{\gamma
-1}}{m(\gamma-1)}\left(\frac{r_{c}^5f(r_c)_{,r_c}}{4}\right)^{(\frac{\gamma-1}{2})}=constant, 
\end{equation}
along with the critical points
\begin{equation}\label{P10}
   (\gamma -1-v_c^2)(\frac{1-v_c^2}{r_c^4
f(r_c)v_c^2})^{\frac{\gamma-1}{2}}=\frac{n_c}{2L}(r_c^5f(r_c)r_c)^\frac{1}{2}v_c^2,
\end{equation}
\begin{equation}\label{P11}
    v_c^2=\frac{r_c f(r_c),r_c}{r_cf(r_c),r_c+4f(r_c)}.
\end{equation}

\begin{figure*}
  \centering
   \includegraphics[scale=0.39]{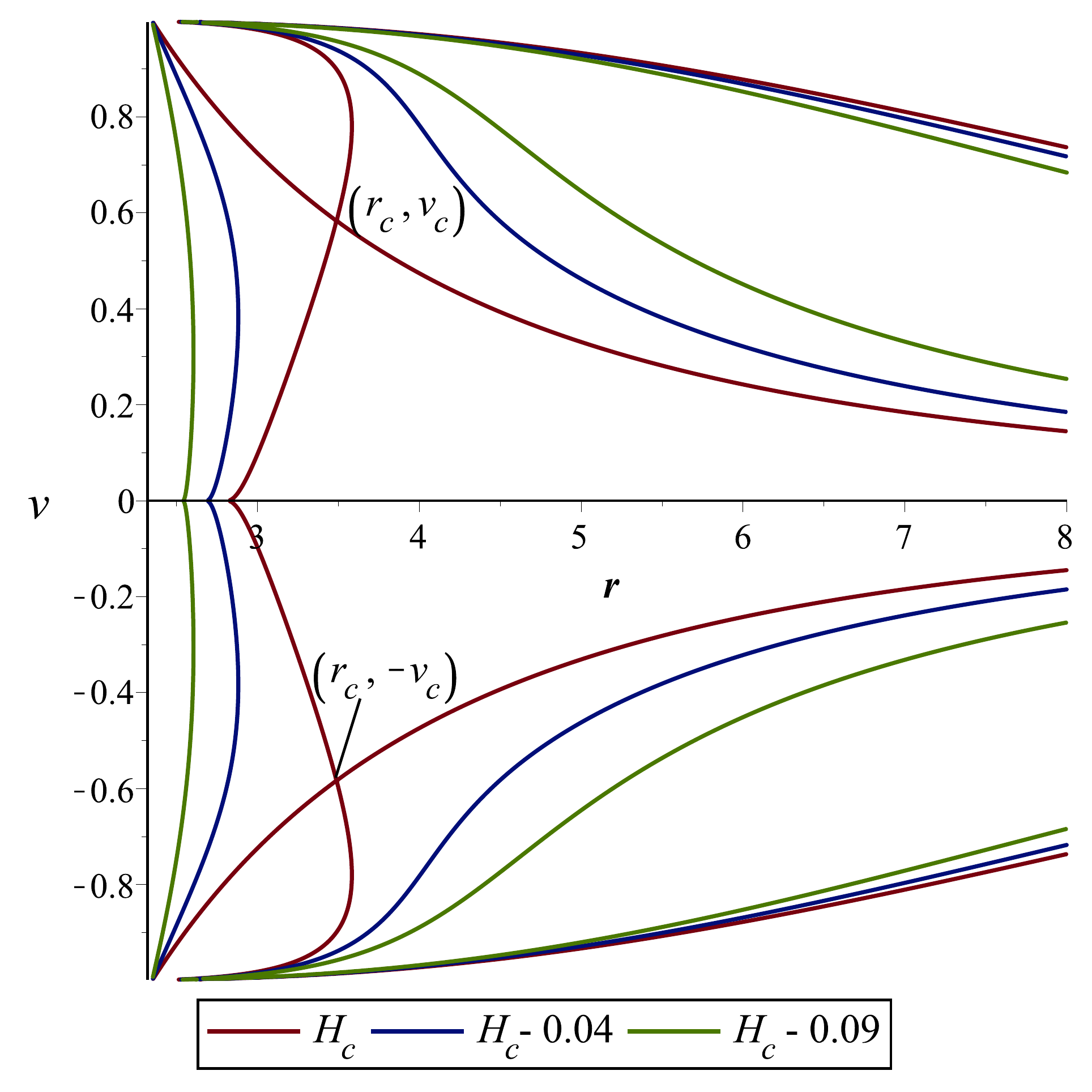}
   \includegraphics[scale=0.39]{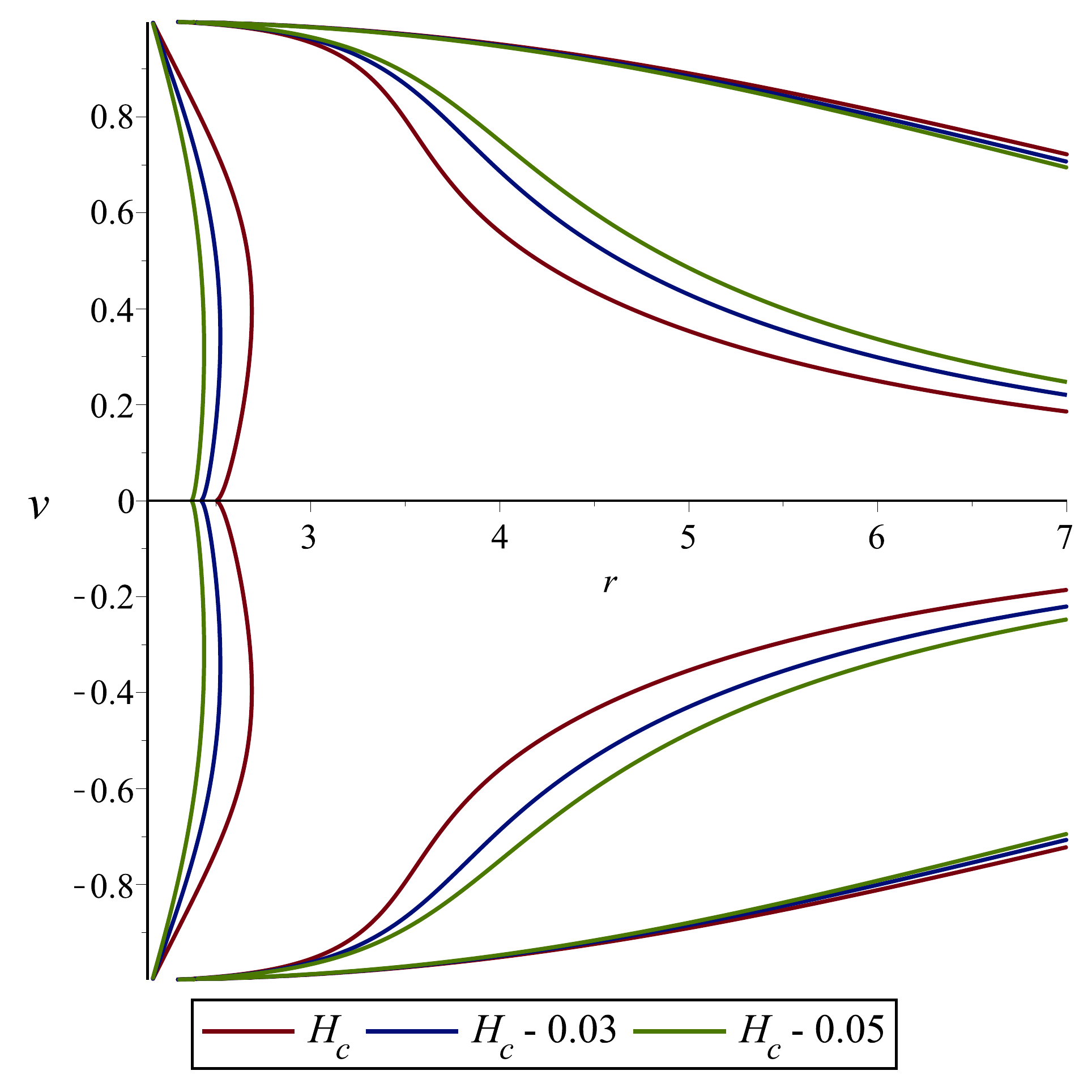}
   \caption {Left panel is the contour plot for infalling gas model seen by distant observer using $S=0.35$($\lambda=1$) with $M=1,  n_c = 0.2$,$\gamma=0.535$ and $ L =
-0.2 $. The parameters are $r_c\approx 3.480626053, v_c \approx
0.5847583391, H_c \approx 0.1650133794$. Right panel is the contour
plot for infalling gas model seen by distant observer using $S=0.35$($\lambda=2$) with $M=1, , n_c = 0.2$,$\gamma=0.5$ and $ L =
-0.2 $. The parameters are $r_c\approx 2.297206433, v_c \approx
0.8939292467, H_c \approx 0.1429485501$.}
\end{figure*}

\begin{figure*}
  \centering
   \includegraphics[scale=0.39]{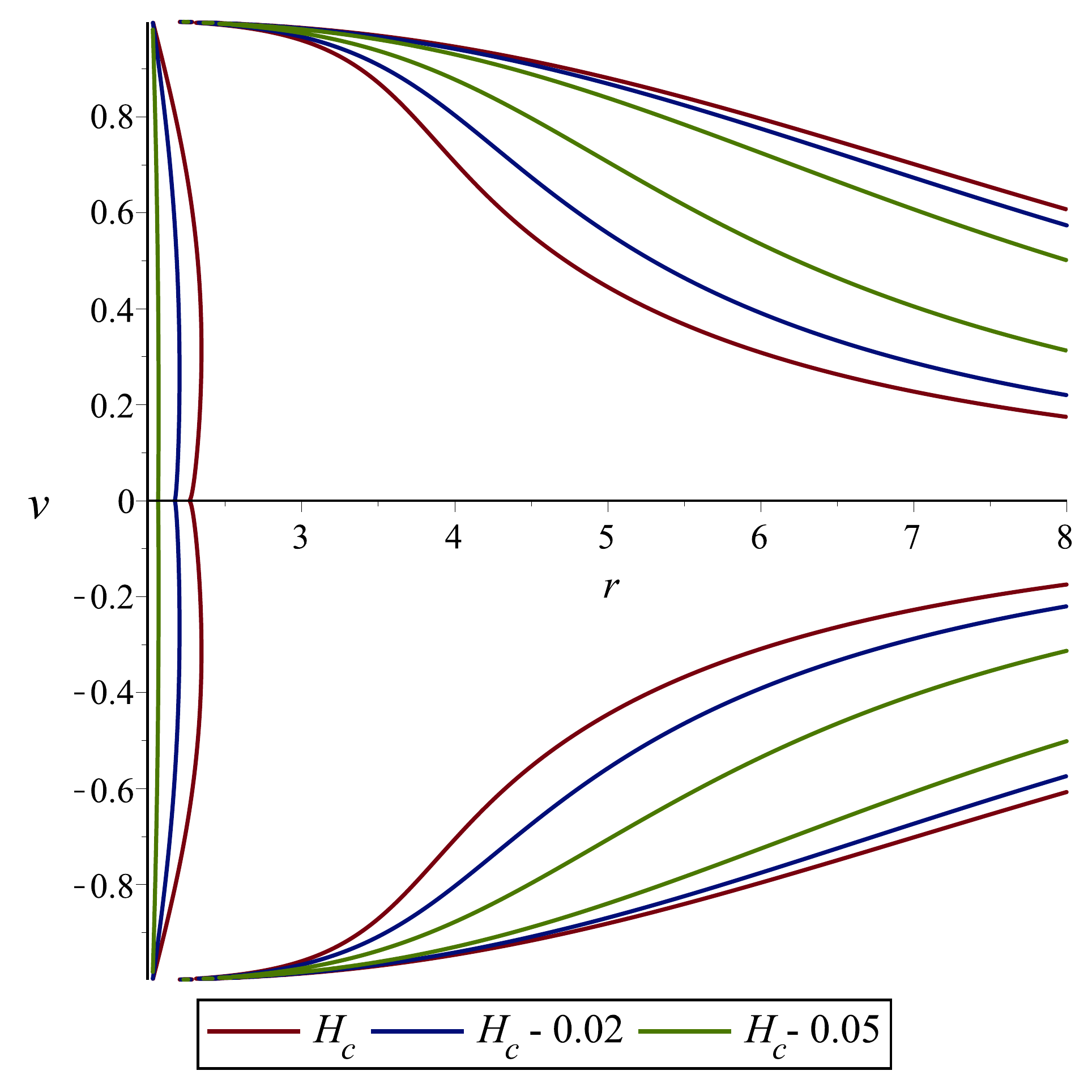}
      \includegraphics[scale=0.39]{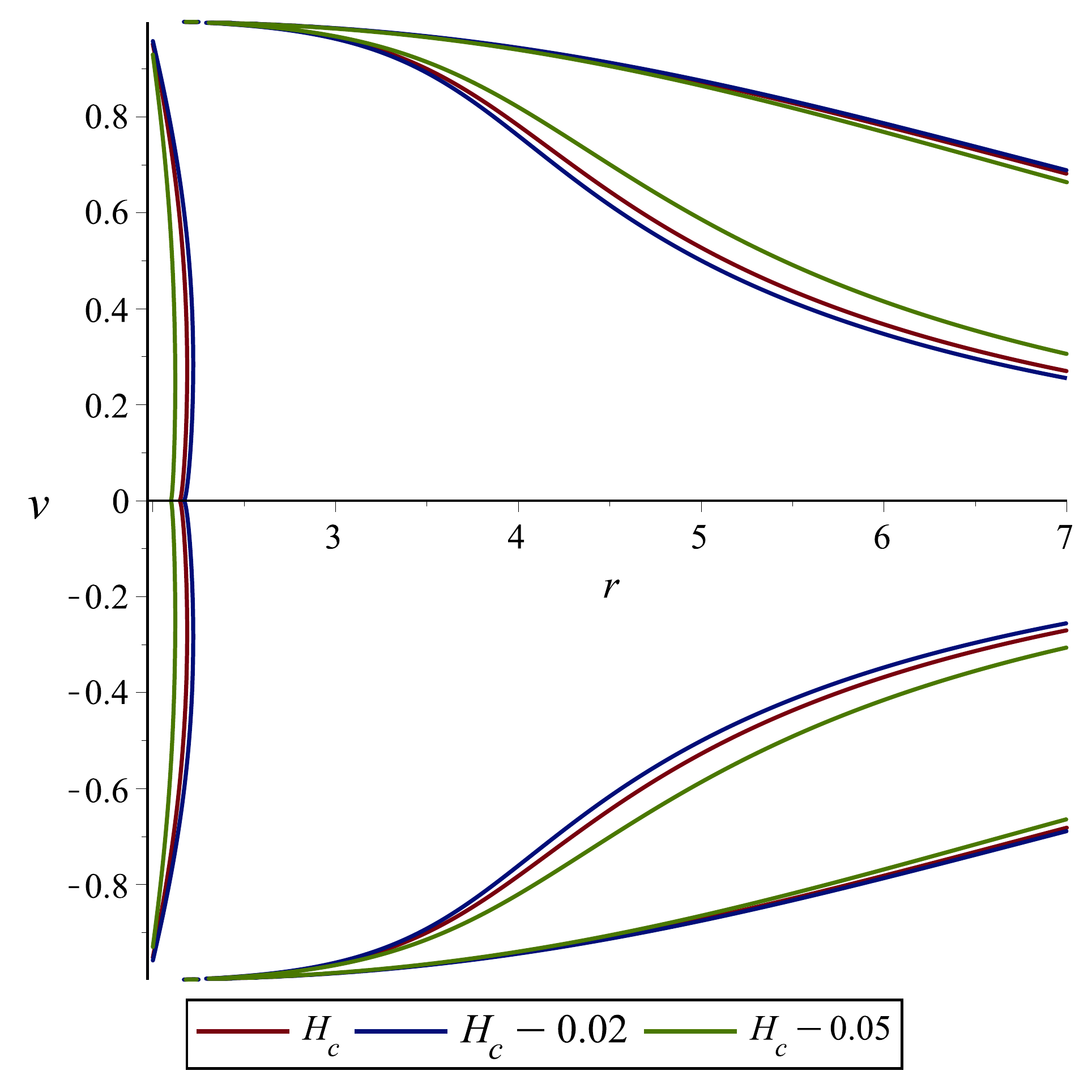}
   \caption {Left panel is the contour plot for infalling gas model seen by distant observer using $S=0.1$($\lambda=3$) with $M=1,  n_c = 0.2$,$\gamma=0.5$ and $ L =
-0.2 $. The parameters are $r_c\approx 1.894211057, v_c \approx
1.159808543, H_c \approx 0.10765175091$. Right panel is the contour
plot for infalling gas model seen by distant observer using $S=-0.1$($\lambda=3$) with $M=1,  n_c = 0.2$,$\gamma=0.5$ and $ L =
-0.2 $. The parameters are $r_c\approx 1.732859099, v_c \approx
1.403403718, H_c \approx 0.07812913389$.}
\end{figure*}

\begin{figure*}
  \centering
   \includegraphics[scale=0.39]{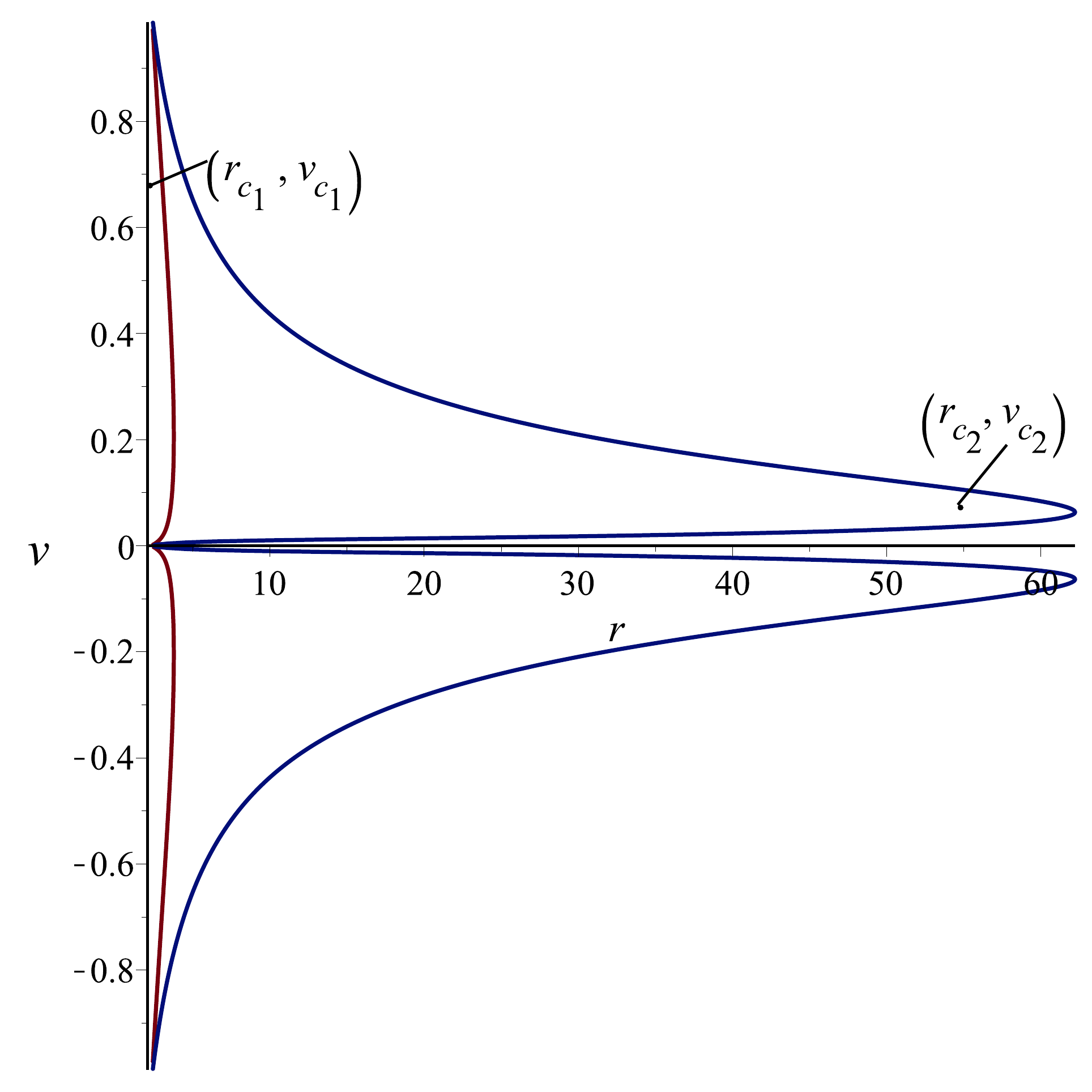}
   \includegraphics[scale=0.39]{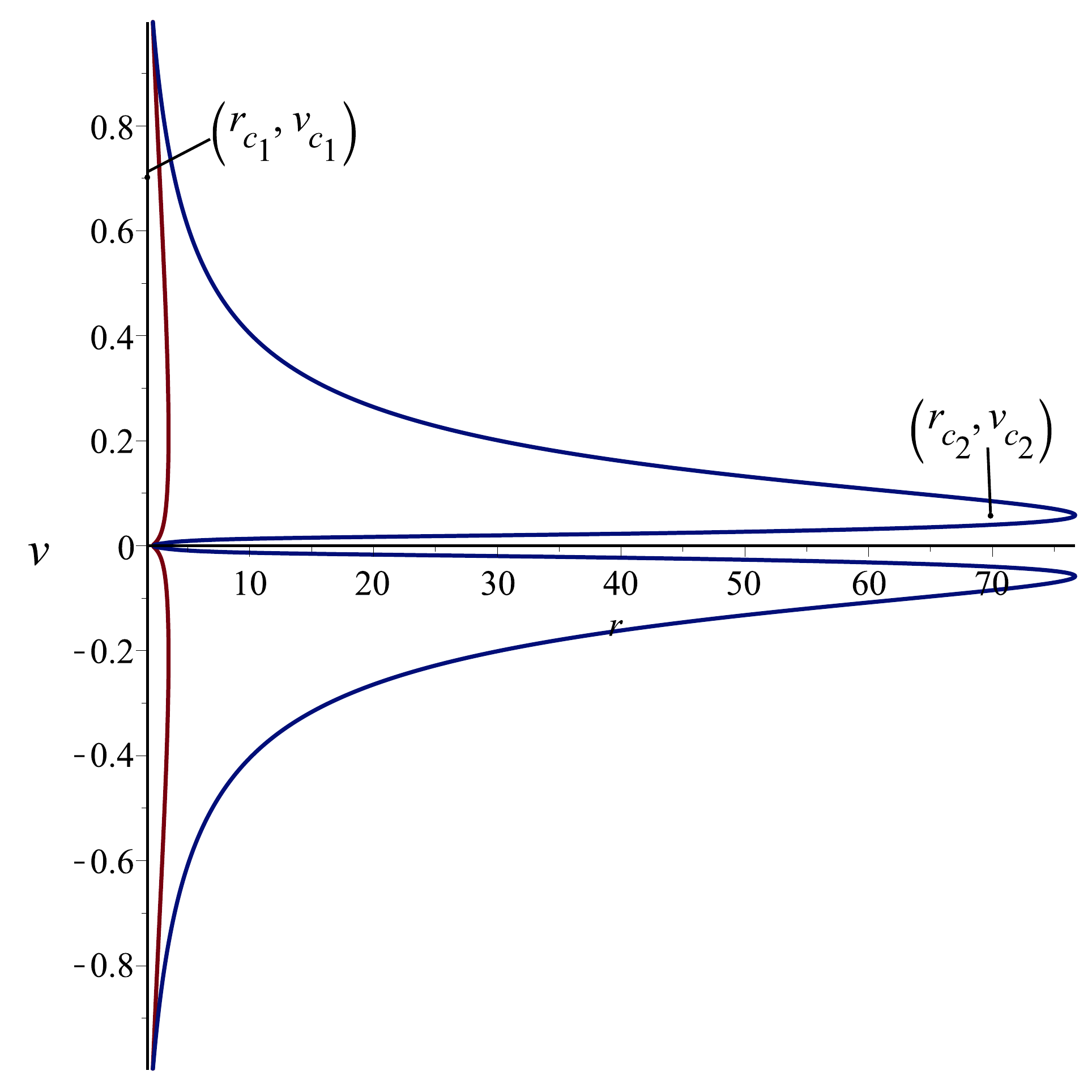}
   \caption {Left panel is the contour plot for infalling gas model seen by distant observer using $S=0.35$($\lambda=1$) with $M=1,  n_c = 0.001$,$\gamma=1.5$ and $ L =
0.125 $. The parameters are $r_{c_1}\approx 3.012941963, v_{c_1} \approx
0.6854442942, H_{c_1} \approx 0.4684637242$ and $r_{c_2}\approx 55.08762356, v_{c_2} \approx
0.1049634266, H_{c_2} \approx 0.9817239945$. Right panel is the contour
plot for infalling gas model seen by distant observer using $S=0.35$($\lambda=2$) with $M=1,  n_c = 0.001$,$\gamma=1.5$ and $ L =
0.125 $. The parameters are $r_{c_1}\approx 2.748090692, v_{c_1} \approx
0.6898671753, H_{c_1} \approx 0.4912239522$ and $r_{c_2}\approx 69.55648888, v_{c_2} \approx
0.08593066630, H_{c_2} \approx 0.9904976826$.}
\end{figure*}

\begin{figure*}
  \centering
   \includegraphics[scale=0.39]{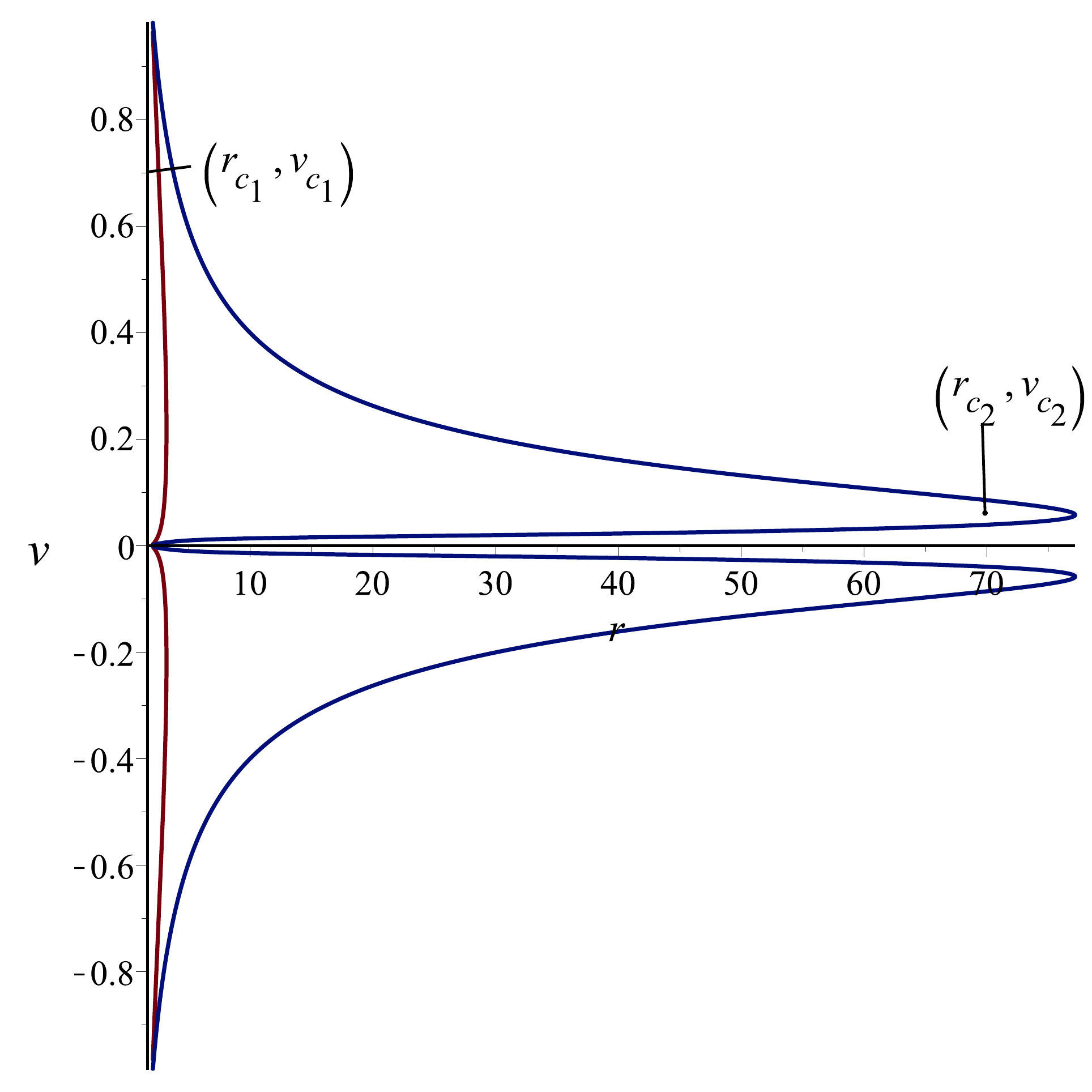}
   \includegraphics[scale=0.39]{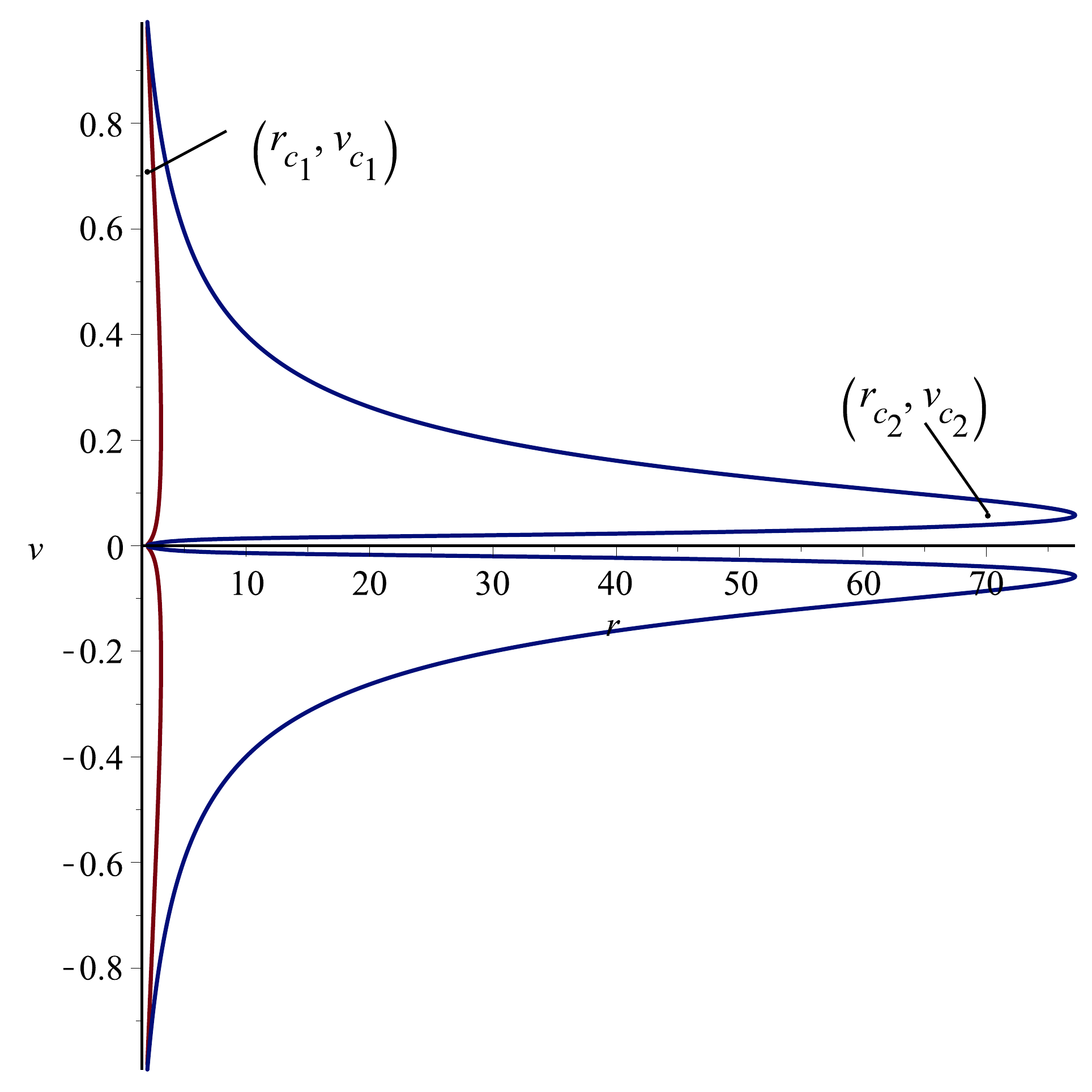}
   \caption {Left panel is the contour plot for infalling gas model seen by distant observer using $S=0.1$($\lambda=3$) with $M=1, n_c = 0.0.001$,$\gamma=1.5$ and $ L =
0.125 $. The parameters are $r_{c_1}\approx 2.567839201, v_{c_1} \approx
0.6932484363, H_{c_1} \approx 0.4769661820$ and $r_{c_2}\approx 70.06543962, v_{c_2} \approx
0.08539629339, H_{c_2} \approx 0.9906416885$. Right panel is the contour
plot for infalling gas model seen by distant observer using $S=-0.1$($\lambda=3$) with $M=1,  n_c = 0.001$,$\gamma=1.5$ and $ L =
0.125$. The parameters are $r_{c_1}\approx 2.508290694, v_{c_1} \approx
0.6943815761, H_{c_1} \approx 0.4664014420$ and $r_{c_2}\approx 70.07167753, v_{c_2} \approx
0.08538979385, H_{c_2} \approx 0.9906431420$.}
\end{figure*}

Figure $(3)$ shows the contour plot for the corresponding intensities using infalling gas model seen by distant observer using $S=0.35 (\lambda =1)$ (left panel) and
using $S=0.35 (\lambda =2)$ (right panel).  We observe that
the accretion for corresponding intensities for $S=0.35 (\lambda =2)$ begins
from subsonic flow for $r$ approaches to infinity, then follows
supersonic flow avoiding the saddle (critical) point and ends into
the horizon. The supersonic outflow starts in the region of horizon
avoiding the saddle (critical) point and ends to subsonic flow for $r$
approaches to infinity. On the other hand, the accretion for corresponding intensities for $S=0.35 (\lambda =1)$ begins
from subsonic flow for $r$ approaches to infinity, then follows
supersonic flow passes through critical (saddle) point and ends into
the horizon. The supersonic outflow starts in the region of horizon
passing the critical (saddle) point and ends subsonically for $r$
approaches to infinity. Figure $(4)$ shows the contour plot for the corresponding intensities using infalling gas model seen by distant observer using $S=0.1 (\lambda =3)$ (left panel) and
using $S=-0.1 (\lambda =3)$ (right panel). We analyze that
the accretion for both corresponding intensities  starts
from subsonic flow for $r$ approaches to infinity, then follows
supersonic flow avoiding the saddle (critical) point and ends into
the horizon. The supersonic outflow starts in the region of horizon
avoiding the saddle (critical) point and ends subsonically for $r$
approaches to infinity.
In Figure (5) and (6), we present the contour plot for different corresponding intensities using infalling gas model seen by distant observer. We presented three sorts of fluid flow, $(i)$ subsonic
non-global flow, $(ii)$non-heteroclinic flow, $ (iii)$ non-relativistic outflow because the solution
curves avoiding the critical points and accretion starts from the
furthest left point until the horizon. 

\section{Black Hole's Accretion Rate}
Mass accretion rate is described by the change in mass of BH per unit time. It is the area times flux of BH
at the event horizon and symbolically represented by \.{M}. Here, we will  point out factor that how the radius of accretion changes by the effect of different fluids. The relativistic statement of the flux of mass-energy density is modeled by $\dot{M}|_{rh}=4 \pi r^2 T_t^r |_{rh}$ (see, \cite{21}-\cite{34}). The energy-momentum tensor of perfect fluid is given by $T_t^r=(\rho+p)u_t
u^r$ \cite{17,24}. As energy of our dynamical system is conserved so
we have $\nabla_\mu J^\mu \equiv 0$ and $\nabla_\nu T^{\mu\nu}
\equiv 0$, from these conservation equations it follows that
\begin{equation}\label{m1}
    r^2 u^r (\rho+p) \sqrt{(f(r)+(u^r)^2)} =K_3,
\end{equation}
where $K_3$ is a constant of integration. Now considering the equation of state $p
\equiv p(\rho)$, the equation of relativistic energy flux is  (\cite{21}-\cite{34})
\begin{equation}\label{m2}
    \frac{d\rho}{\rho+p}+\frac{du^r}{u^r}+2\frac{dr}{r} = 0.
\end{equation}
By integrating the above equation,
\begin{equation}\label{m3}
    r^2 u^r \exp\Big[\int_{\rho \infty}^\rho \frac{d\acute{\rho}}{\acute{\rho}+p(\acute{\rho})}\Big] =
    -K_4,
\end{equation}
where $K_4$ is a constant of integration. After simplify the equation (\ref{m3}) and
(\ref{m1}), it can be shown
\begin{equation}\label{m4}
    K_5=-\frac{K_3}{K_4}=(\rho+p)\sqrt{Y(r)+(u^r)^2}
    \exp\Big[-\int\frac{d\acute{\rho}}{\acute{\rho}+p(\acute{\rho})}\Big] ,
\end{equation}
where $K_5$ is a constant. Consider the boundary condition at
infinity, then $K_5 \equiv
\rho_\infty+p(\rho_\infty)=-\frac{K_3}{K_4}$, here $K_3 \equiv
(\rho+p)u_t u^r r^2 = -K_4(\rho_\infty + p(\rho_\infty))$. Using above relations it follows that
\begin{equation}\label{m6}
    \frac{\rho+p}{n} \sqrt{f(r)+(u^r)^2} \equiv K_7,
\end{equation}
where $K_7$ is a constant such that
$K_7=\frac{(\rho_\infty+p_\infty)}{n_\infty}$. Using (\ref{m1}), we
get new relation of BH's accretion rate
\begin{equation}\label{m7}
    \dot{M}=-4 \pi r^2 u^r (\rho + p)\sqrt{f(r)+(u^r)^2} = -4 \pi
    K_3,
\end{equation}
then it becomes
\begin{equation}\label{m8}
\dot{M}=-4 \pi K_4(\rho_\infty+p(\rho_\infty)).
\end{equation}
The above result holds for all those fluids due to boundary condition at infinity on which the equation of state holds in
the form $p = p(\rho)$. So, the accretion rate of BH takes the form
\begin{equation}\label{m9}
\dot{M}=-4 \pi K_4(\rho+p)|_{r=r_h},
\end{equation}
at event horizon $r_h$. 

Let us take an isothermal equation of state, i.e., $p=k \rho$, which implies that
$(\rho + p) \equiv \rho (1+k)$. Then equation (\ref{m3}) reduces to
$r^2 u^r \rho^{\frac{1}{1+k}}= - K_4$, that is
\begin{equation}\label{m10}
    \rho = \Big[-\frac{K_4}{r^2 u^r}\Big]^{1+k} ,
\end{equation}
using above information, equation (\ref{m1}) takes the form
\begin{equation}\label{m11}
    (u^r)^2 - \frac{K_3^2 K_4^{-2(1+k)}}{(1+k)^2} r^{4k} (-u^r)^{2k}
    + f(r)=0,
\end{equation}
in which the $u^r$ can be obtained for the given values of $k$.
Using $u^r$, one can find energy density $\rho(r)$ from (\ref{m10}).

\begin{figure}
  \centering
   \includegraphics[width=7.5cm]{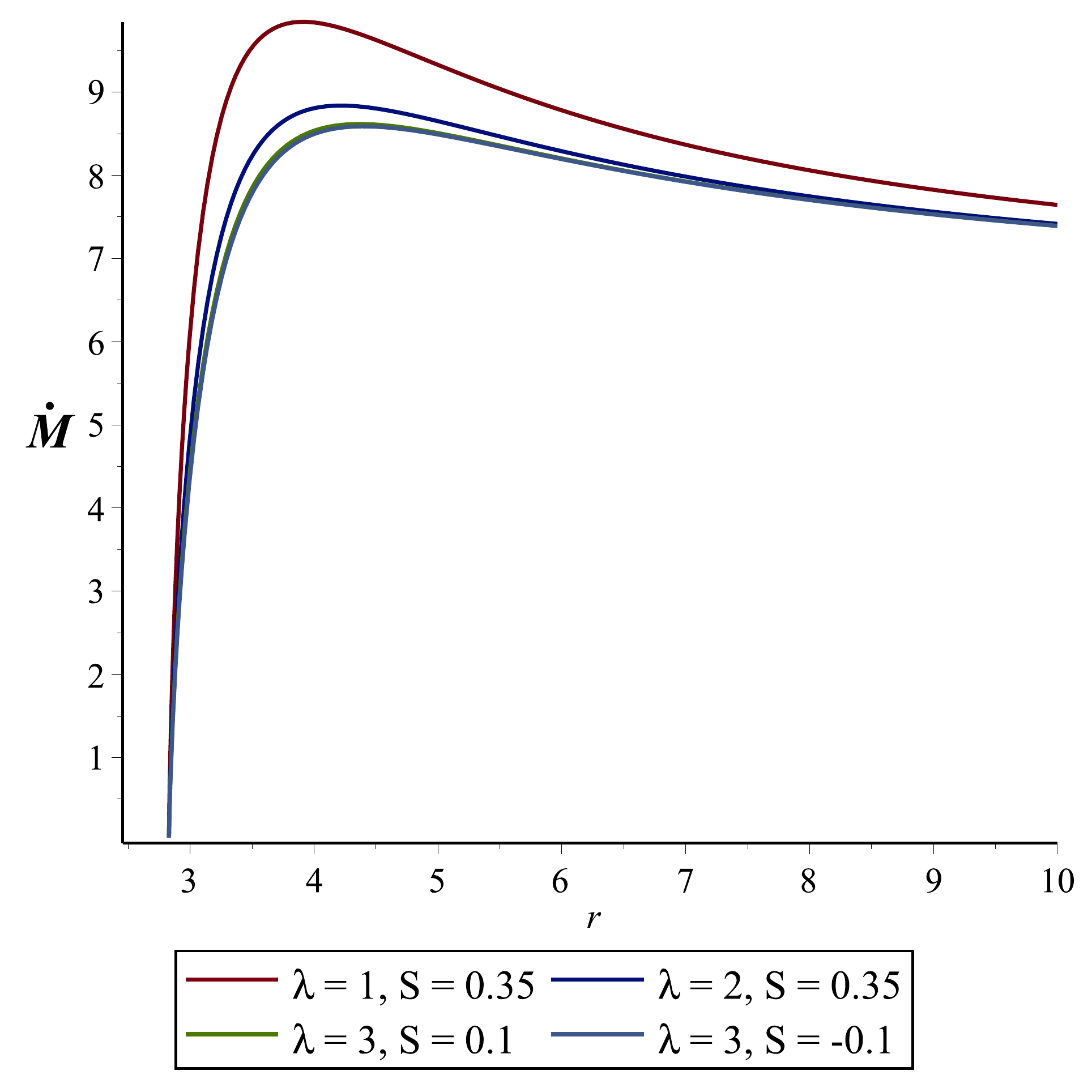}
   \caption {Contour plot of mass accretion rate for different intensities of infalling gas model seen by distant observer with $M=1, K_3=1, K_4=2$}
\end{figure}

\subsection{Behavior of ultra-stiff fluid} 
 The four-velocity $u^r$ is obtained by taking $k=1$ in
(\ref{m11}), that is \cite{34a,34b,34c}
\begin{equation}\label{m12}
    u^r = \pm 2K_4 \sqrt{\frac{f(r)}{K_3^2 r^4- 4K_4^4}} .
\end{equation}
The corresponding energy density is 
\begin{equation}\label{m13}
    \rho = \frac{(K_3^2 r^4-4K_4^4)}{4 K_4^2 r^4 f(r)}.
\end{equation}
One can find the mass accretion
rate by using (\ref{m13}) and (\ref{m9})
\begin{equation}\label{m14}
\dot{M}=\frac{2 \pi (K_3^2 r^4-4K_4^4)}{K_4 r^4 f(r)}.
\end{equation}
Fig. (7) represent the behavior of mass accretion rate for different intensities of infalling gas model seen by distant observer  for the ultra-stiff
fluid $(k=1)$. One can notice that mass accretion rate
depends upon the model parameter $\lambda$ which further depends upon the scalar charge $S$. The mass accretion rate is increased for higher values of scalar charge $S$. We find the
maximum accretion rate for 
\begin{itemize}
  \item $\lambda=1, S=0.35$, we have $r\approx 3.9$.
  \item $\lambda=2, S=0.35$, we have $r\approx 4.0$.
  \item $\lambda=3, S=0.1$, we have $r\approx 4.1$.
  \item $\lambda=3, S=-0.1$, we have $r\approx 4.12$.
\end{itemize}
So, we can conclude that mass accretion rate depend upon the
parameters $(\lambda, S)$. It is interesting to mention here,
that mass accretion rate is higher for lower values of scalar charge.

\subsection{Behavior of ultra-relativistic fluid }

The four-velocity $u^r$ is obtained by taking
$k=\frac{1}{2}$ in (\ref{m11}), that is \cite{34a,34b,34c}
\begin{equation}\label{k1by2ur}
    u^r = \frac{2 r^2 K_3^2+ \sqrt{4 r^2 K_3^4 - 81 f(r) K_4^6}}{9
    K_4^3},
\end{equation}
and corresponding energy density is given by
\begin{equation}\label{m15}
    \rho =27\Big(\frac{K_4^4}{r^2(2r^2 K_3^2+ \sqrt{4 r^4K_3^4-81 f(r)
    K_4^6 })}\Big)^{3/2}.
    \end{equation}
By using (\ref{m15}) in (\ref{m9}), we get
\begin{equation}\label{m16}
\dot{M}=216 \pi K_4 \Big[\frac{K_4^4}{r^2(2 r^2 K_3^2 + \sqrt{4 r^4
K_3^4 -81 f(r)K_4^6})} \Big]^{3/2}.
\end{equation}

Fig. (8) represent the behavior of mass accretion rate for different intensities for $(k=\frac{1}{2})$ (ultra-relativistic fluid). We can
find the maximum accretion rate for 
\begin{itemize}
  \item $\lambda=1, S=0.35$, we have $r_{IH}\approx0.1, r_{OH}\approx2.336$
  \item $\lambda=2, S=0.35$, we have  $r_{IH}\approx0.15, r_{OH}\approx 2.15$
  \item $\lambda=3, S=0.1$, we have $r_{IH}\approx0.07, r_{OH}\approx2$
  \item $\lambda=3, S=-0.1$, we have $r_{IH}\approx0.25, r_{OH}\approx 1.96$
\end{itemize}
From above values, it is evident that the maximum mass accretion rate located at inner horizon (IH) and outer horizon (OH). So, we can conclude that mass accretion rate depend upon the
constants $(\lambda,S)$ and its maximum accretion rate occur at
$\lambda = 3$ and $S=-0.1$ as compare to other values.
\begin{figure}
  \centering
   \includegraphics[width=8cm]{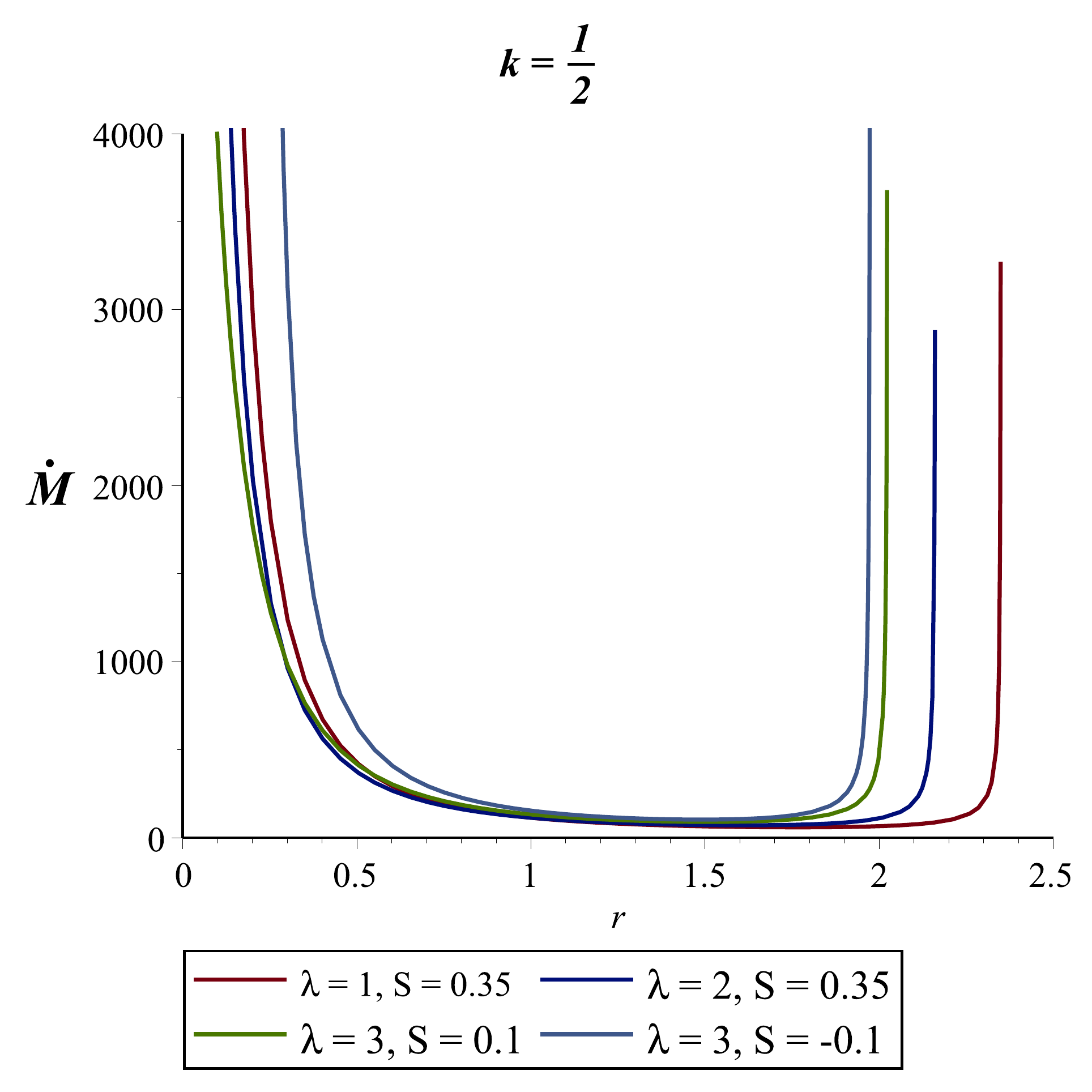}
   \caption {Contour plot of mass accretion rate for different intensities with $M=1,
   K_3=0.1, K_4=2$.}
\end{figure}

\subsection{Behavior of radiation fluid}
The radial velocity for $k=\frac{1}{3}$ is given by \cite{34a,34b,34c} 
\begin{widetext}
\begin{eqnarray}\nonumber
    u^r &=& \Big[ \frac{\big(-32 f(r) K_4^4 + \sqrt{1024 f(r)^2 K_4^8 -27 r^4 K_3^6} K_4^2\big)^{1/3}}{4
    K_4^2}\\
    &+& \frac{3 r^{4/3} K_3^2}{4 K_4^{2/3}\big((-32 f(r) K_4^4 + \sqrt{1024 f(r)^2 K_4^8 -27 r^4
    K_3^6})
    K_4^2\big)^{1/3}}\Big]^{2/3}\label{k1by3ur}.
\end{eqnarray}
The energy density
put the value of $u^r$ as calculated above and  $k=\frac{1}{3}$ in (\ref{m10}), we get
\begin{eqnarray}\nonumber
    \rho &=&\Big[\frac{K_4}{r^2}\Big]^{\frac{4}{3}} \Big[ \frac{\big(-32 f(r) K_4^4 + \sqrt{1024 f(r)^2 K_4^8 -27 r^4 K_3^6} K_4^2\big)^{1/3}}{4
    K_4^2}\\
    &+& \frac{3 r^{4/3} K_3^2}{4 K_4^{2/3}\big((-32 f(r) K_4^4 + \sqrt{1024 f(r)^2 K_4^8 -27 r^4
    K_3^6})
    K_4^2\big)^{1/3}}\Big]^{\frac{-8}{9}}\label{m17}.
    \end{eqnarray}

By using (\ref{m17}) and (\ref{m9}), we can get mass accretion rate
\begin{eqnarray}\nonumber
\dot{M}&=&\Big[\frac{8 \pi
K_4^{\frac{7}{3}}}{r^{\frac{8}{3}}}\Big]\Big[\big(-32 f(r)
K_4^4+ \sqrt{1024 f(r)^2 K_4^8 -27 r^4 K_3^6} K_4^2\big)^{1/3}(4K_4^2)^{-1} \\
&+& \frac{3 r^{4/3} K_3^2}{4 K_4^{2/3}}\Big((-32 f(r) K_4^4 + \sqrt{1024 f(r)^2 K_4^8 -27 r^4
    K_3^6})
    K_4^2\big)^{-1/3}]^{\frac{-8}{9}}\label{m18},
\end{eqnarray}
\end{widetext}

Fig. (9) represent the behavior of mass accretion rate for different intensities for
radiation fluid. Maximum mass accretion rate is located at inner and outer horizons for radiation fluid.
So, we can conclude that mass accretion rate depend upon the
constants $(\lambda,S)$ and its maximum accretion rate occur at
$\lambda =3$ and $S=0.1$ as compared to other values.
\begin{figure}
\centering
  \includegraphics[width=8 cm]{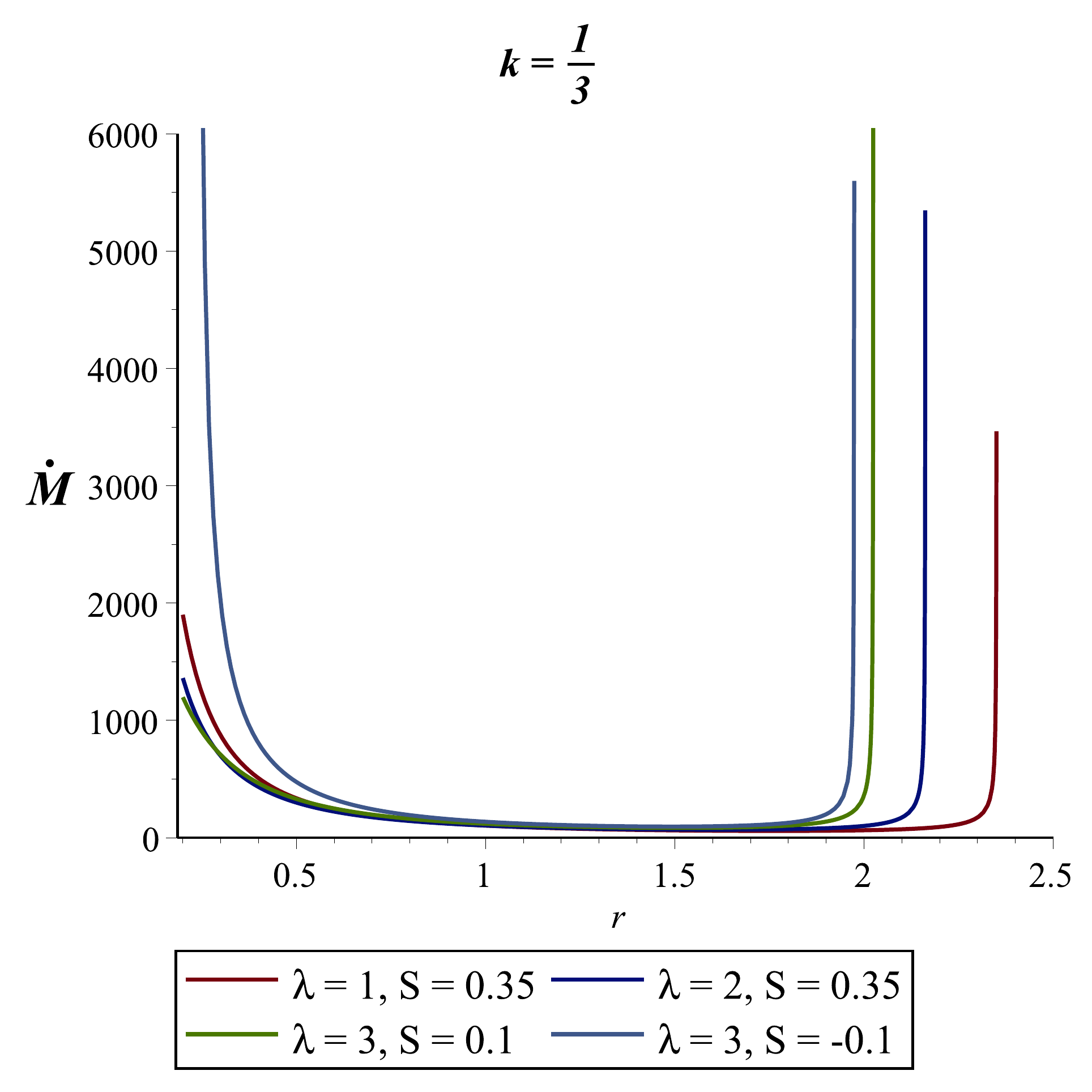}
   \caption {Contour plot of mass accretion rate for  different intensities with $M=1,
   K_3=0.1, K_4=2$.}
\end{figure}

\subsection{Behavior of sub-relativistic fluid}
The mass accretion rate for different corresponding intensities of infalling gas model seen by distant observer
is shown in Fig. $(10)$ for
sub-relativistic fluid. One can notice that mass
accretion rate increase for higher values of lower values of scalar charge i.e. $\lambda=2, S=-0.45$ as compare to other values.
One can find max accretion rate near \begin{itemize}
  \item $\lambda=1, S=0.35$, we have $r_{IH}\approx0.05, r_{OH}\approx2.34$
  \item $\lambda=2, S=0.35$, we have  $r_{IH}\approx0.05, r_{OH}\approx2.15$
  \item $\lambda=3, S=0.1$, we have $r_{IH}\approx0.05, r_{OH}\approx2$
  \item $\lambda=3, S=-0.1$, we have $r_{IH}\approx0.2, r_{OH}\approx 1.95$
\end{itemize}
It is evident that the mass accretion rate is infinite near the singularity \cite{34a,34b,34c}.
\begin{figure}
  \centering
   \includegraphics[width=8 cm]{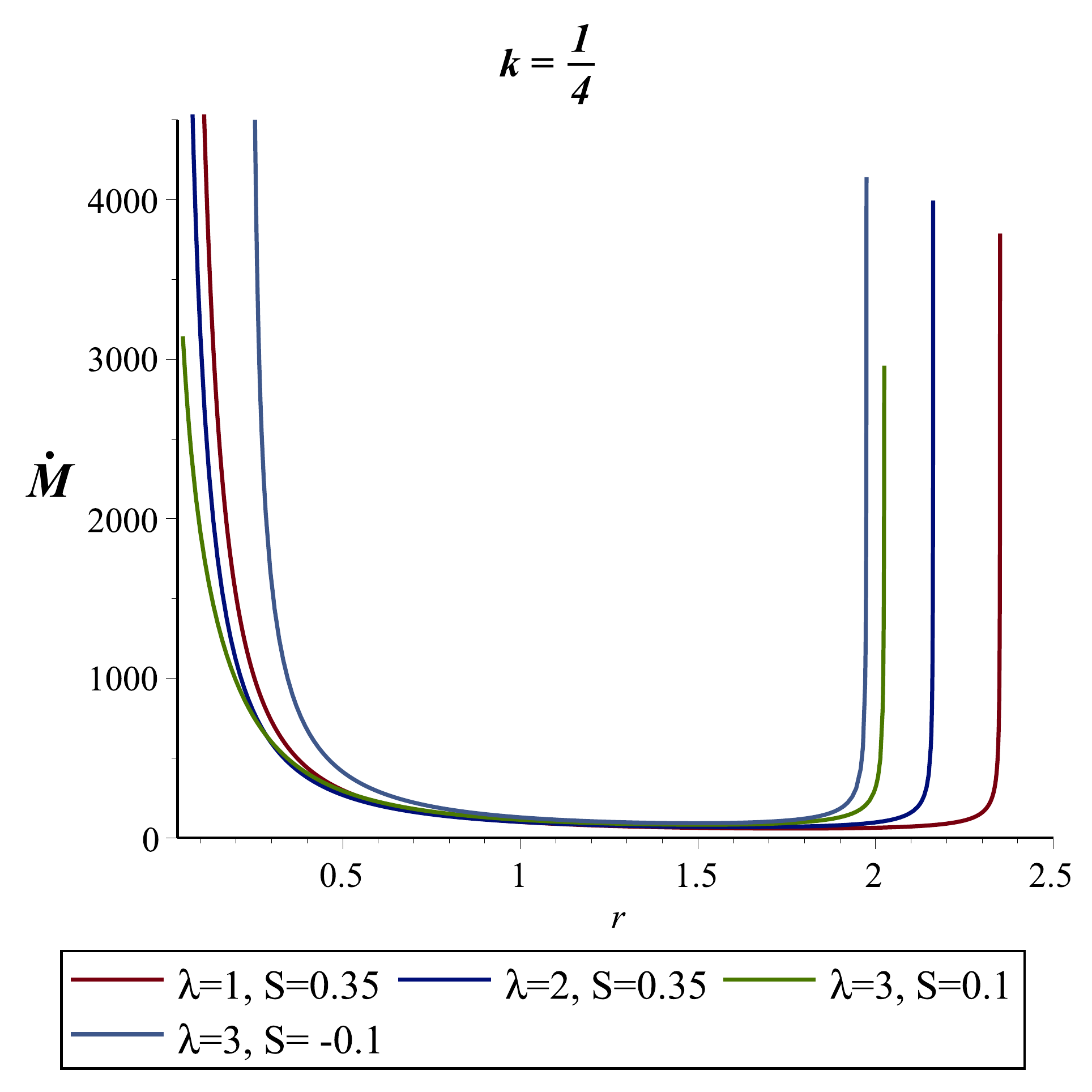}
      \caption {Contour plot of mass accretion rate for  different intensities with $M=1,
      K_3=0.1, K_4=2$.}
\end{figure}

\section{Phase transition via black hole shadows }
In this last section we shall explore an interesting idea to analyse the phase transitions of the black hole based on the black hole shadow formalism developed in \cite{Zhang:2019glo}. 
We know that the specific heat of a black hole can be: positive, i.e., $C>0$ and negative i.e., $C<0$, meaning a thermodynamically stable and unstable state, respectively. In addition when $C=0$ is the phase transition point. Furthermore the entropy $S$ is related to the event horizon $r_+$ of the black hole with
$dS/dr_+>0$, as a result we can write
\begin{equation}\label{sig}
\text{Sgn}(C)= \text{Sgn}\left(\frac{\partial T}{\partial r_+}\right),
\end{equation}
where Sgn is the sign function. In other words, we can say that in the e $r_+ -T$ diagram a positive slope means that the black hole is in a thermodynamically stable state and a negative slope corresponds to a thermodynamically unstable black hole. As shown in Ref. \cite{Zhang:2019glo} and subsequently \cite{Belhaj:2020nqy}, there is a possibility to probe the phase transitions using the shadow radius of the black holes. 
To see this, recall that the effective potential of the photon can be defined by
\begin{equation}
V_{eff}(r)+\dot{r}^{2}=0,
\end{equation}
along with the further condition which allows to obtain the radius of the photon 
\begin{equation}\label{effpho}
V_{eff}(r)=V^{\prime}_{eff}(r)=0.
\end{equation}

\begin{figure*}
    \centering
    \includegraphics[scale=0.6]{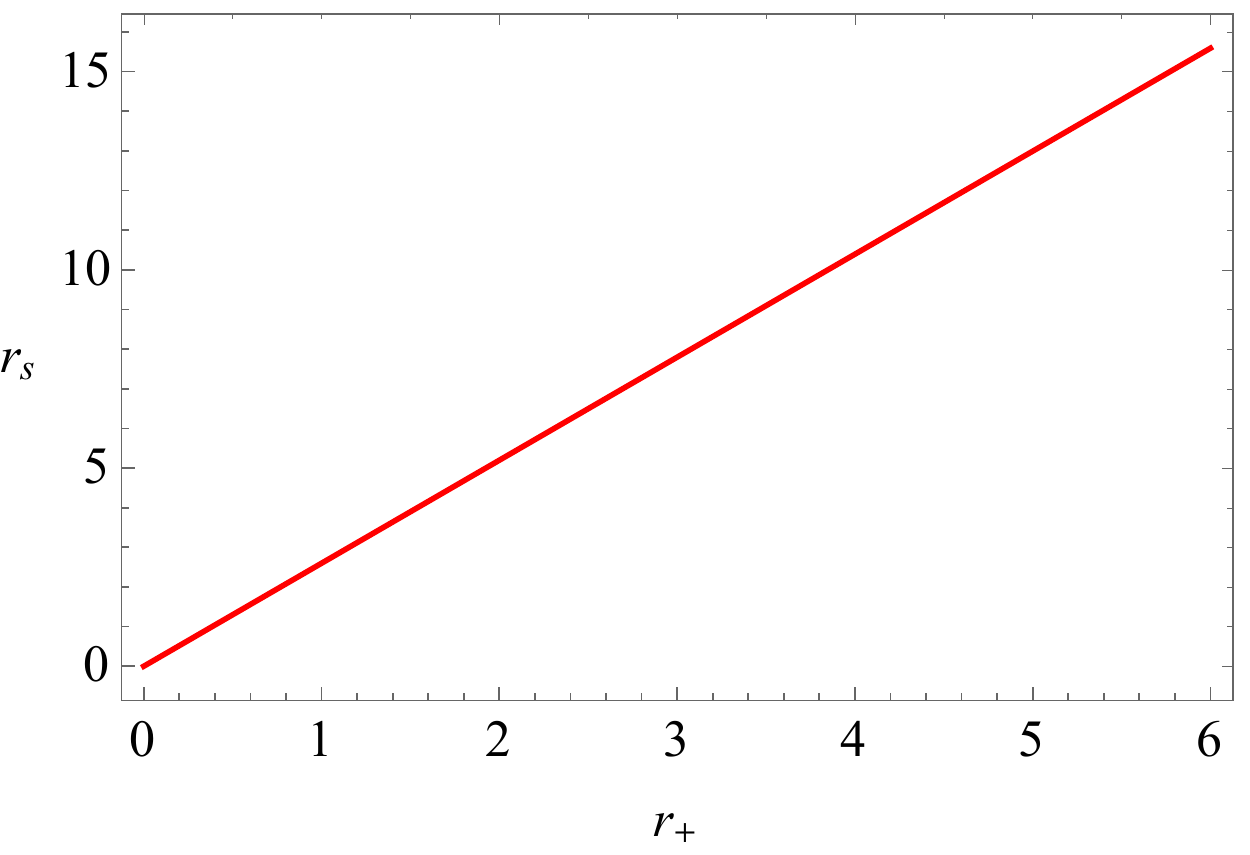}
        \includegraphics[scale=0.6]{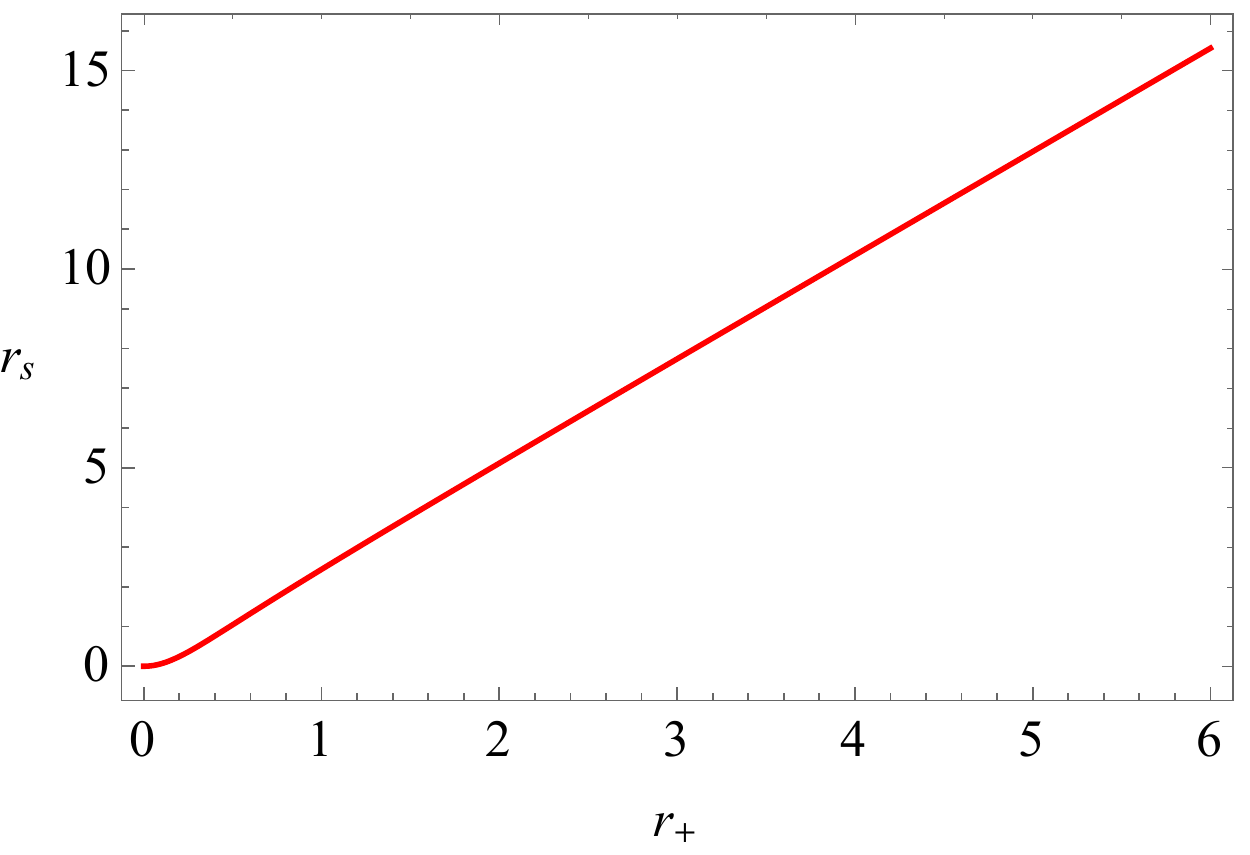}
    \caption{Left panel: Plot of the shadow radius as a function of the event horizon for $\lambda=1$. Right panel:  Plot of the shadow radius as a function of the event horizon for $\lambda=2$ using $S=0.2$. }
\end{figure*}

\begin{figure*}
    \centering
    \includegraphics[scale=0.65]{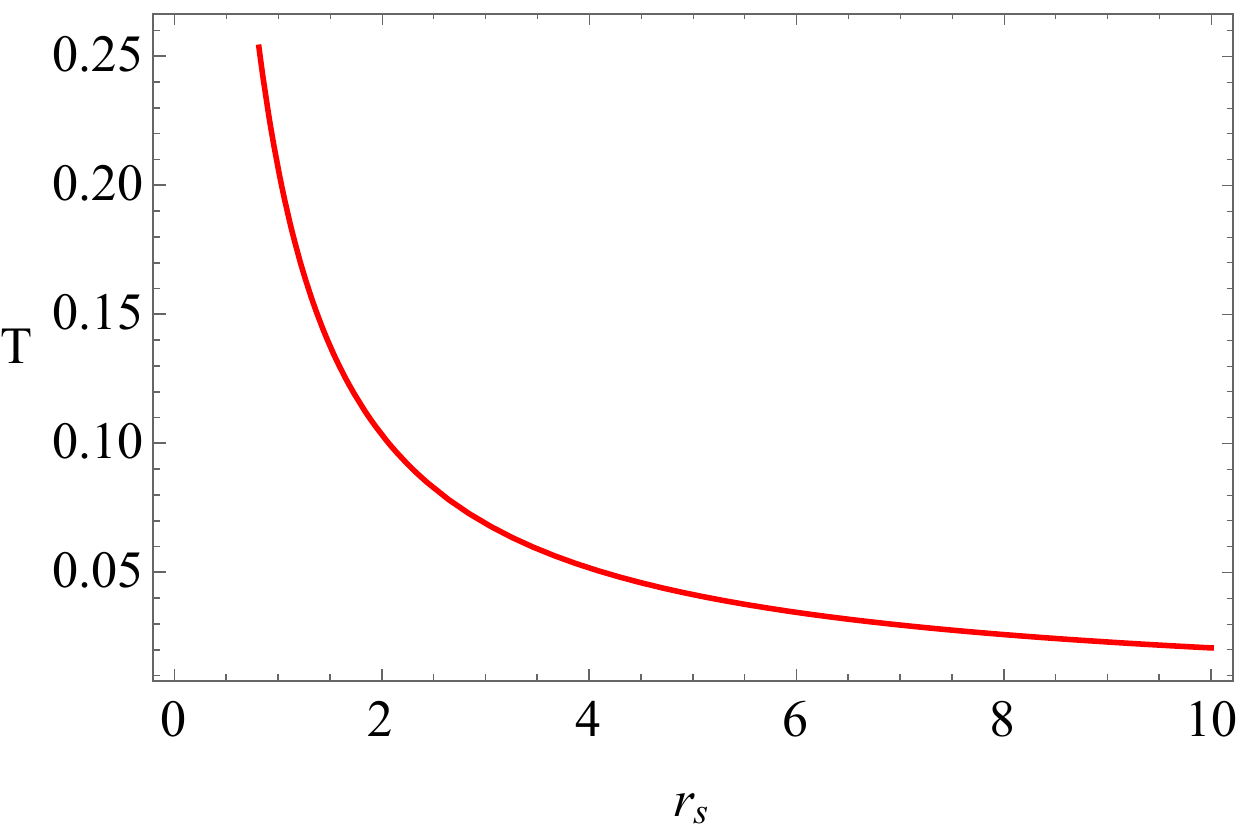}
        \includegraphics[scale=0.65]{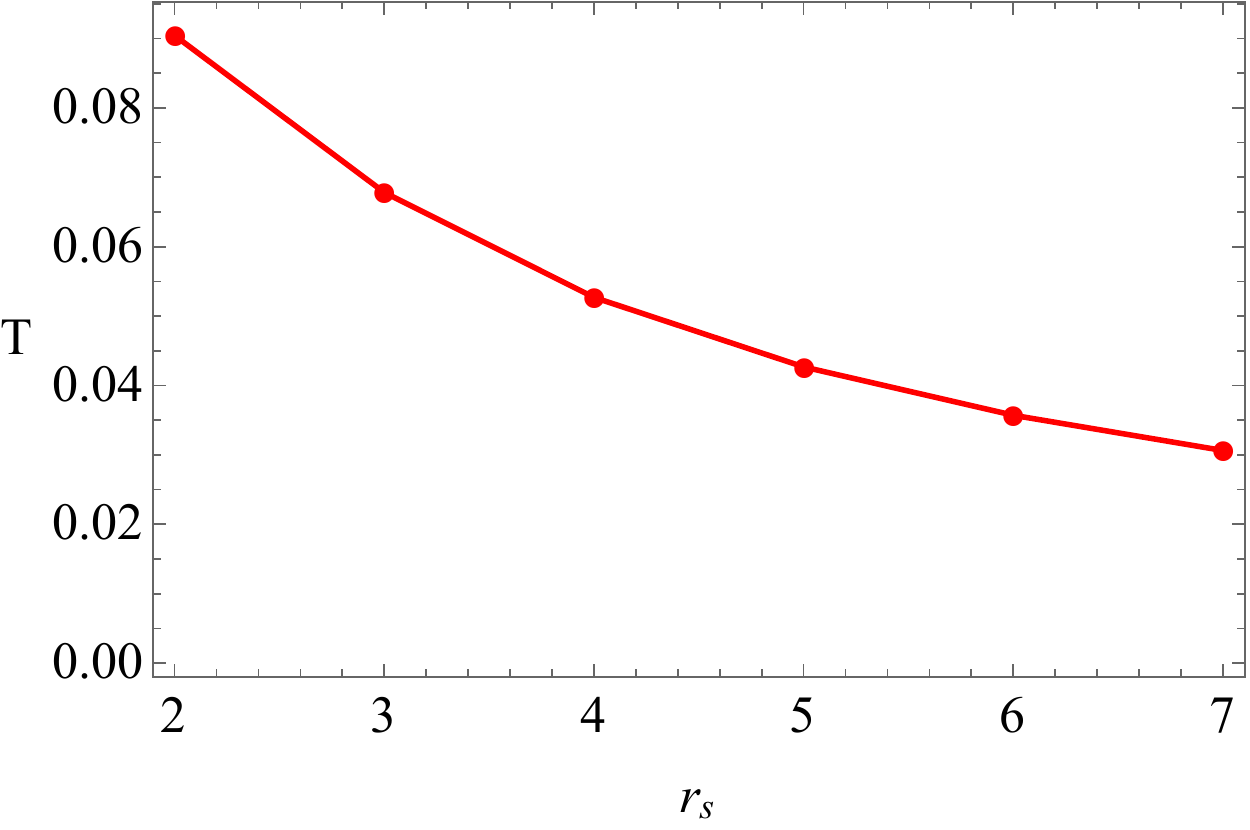}
    \caption{Left panel:  Plot of the black hole temperature as a function of the shadow radius for $\lambda=1$. Right panel:  Plot of the black hole temperature as a function of the shadow radius for $\lambda=2$ and $S=0.2$.  }
\end{figure*}

Considering the circular orbit of the photon with
\begin{equation}
V^{\prime\prime}_{eff}(r)<0,
\end{equation}
and using the fact that the Hawking temperature of the black hole is positive it follows  \cite{Zhang:2019glo}
\begin{equation}\label{dre3}
\frac{dr_s}{dr_+}>0
\end{equation}
Finally for the temperature $T$ of the black hole it was argued that  \cite{Zhang:2019glo}
\begin{equation}
\frac{\partial T}{\partial r_{+}}=\frac{\partial T}{\partial r_{s}}\frac{dr_{s}}{dr_{+}},
\end{equation}
which means
\begin{equation}\label{sphereone}
\frac{\partial T}{\partial r_{+}}>0,~~\frac{\partial T}{\partial r_{+}}=0,~~\frac{\partial T}{\partial r_{+}}<0,
\end{equation}
and
\begin{equation}\label{spheretwo}
\frac{\partial T}{\partial r_{s}}>0,~~\frac{\partial T}{\partial r_{s}}=0,~~\frac{\partial T}{\partial r_{s}}<0,
\end{equation}
respectively. In order to work with analytical expressions we shall simplify our analyses by considering the special case with: $\lambda=1$ and $\lambda=2.$
\subsection{Case I}
In this particular case, $\lambda=1$, we have the following result for the event horizon
\begin{eqnarray}
r_+=2M+S
\end{eqnarray}
along with shadow radius 
\begin{eqnarray}
r_s=\frac{3 \sqrt{3} r_+ }{2},
\end{eqnarray}
and consequently the temperature 
\begin{eqnarray}
T=\frac{1}{4 \pi r_+}.
\end{eqnarray}

As can be observed from Fig. (11) (left panel) and Fig. (12) (left panel), for this particular case, we have
\begin{equation}
\frac{dr_s}{dr_+}>0,\,\,\,\frac{\partial T}{\partial r_{s}}<0,
\end{equation}
meaning that the black hole is thermodynamically unstable.

\subsection{Case II}
Let us take $\lambda=2$, in this case we have the following result for the event horizon
\begin{eqnarray}
r_+=M+\sqrt{M^2+S^2}.
\end{eqnarray}
As a result the shadow radius reads
\begin{equation}
r_s=\frac{(3M+\sqrt{9M^2+8S})}{2\Big[1-\frac{2M}{(3M+\sqrt{9M^2+8S})/2}+\frac{S}{((3M+\sqrt{9M^2+8S})/2)^2}\Big]^{1/2}},
\end{equation}
where 
\begin{eqnarray}
M=\frac{r_+^2-S}{2r_+}.
\end{eqnarray}
Finally, the temperature reads
\begin{eqnarray}
T=\frac{r_+^2-S+r_+S}{4 \pi r_+^3}.
\end{eqnarray}

Similarly, from Fig. (11) and Fig. (12) (right panels)  we see that the slope of the black hole temperature as a function of the shadow radius is negative hence we conclude that the black hole  is thermodynamically unstable for such domain of parameters.

\section{Conclusions}
In this paper we have used a model of infalling and radiating gas and studied the shadow images as well as the intensities in the spacetime of black hole solutions obtained in a theory of massive gravity with a Lorentz violating symmetry. In particular we have shown that the properties of the shadow images depends strongly on the sign before the scalar charge $S$.  In the case $S>0$, we saw that the shadow radii are bigger while the intensities are smaller compared to $S<0$. Our analyses is based on a range of values for $S$ which are obtained by constraining $S$ from the EHT results. 
We discussed the accretion process of different fluids such as sub-relativistic fluid ($k=1/4$), radiation fluid ($k=1/3$), ultra-relativistic fluid ($k=1/2$) and ultra-stiff fluid ($k=1$) on massive gravity black hole in the presence of "Hair parameter" ($\lambda$) and scalar charge $S$. We also investigated the behavior of the polytropic fluid onto massive gravity black hole where the accretion starts from supersonic/subsonic flow and it passes through the saddle (critical) point on the massive gravity black hole using model parameter $\lambda=1$ and scalar charge $S=0.35$ where as it does not pass through the critical point on massive gravity black hole for other parameters. Most importantly, the accretion process ends near the Killing horizon which agrees with other recent studies.

We also explored the four velocity in radial direction ($u^r$), energy density ($\rho$) and mass accretion rate ($\dot{M}$). We investigated the mass accretion rate of massive gravity black hole in the presence of different fluids which indicates important signatures as graphically represented in Figs. 7, 8, 9 and 10.  The mass accretion rate is higher in massive gravity black hole for higher values of scalar charge in the presence of ultra-stiff fluid ($k=1$). The maximum mass accretion rate is located at inner and outer horizons in the presence of radiation and sub-relativistic fluids. For sub-relativistic, radiation and ultra-relativistic fluids accretion in massive black hole, the mass accretion rate is higher for lower values of scalar charge. From figures, it is evident that mass accretion rate is infinite near singularity in the presence of sub-relativistic, radiation and ultra-relativistic fluids.

Our final part of this work is devoted to the problem of phase transition and it's connection to the shadow radius.  We have considered specific cases and found that the black hole can be thermodynamically unstable in a given domain of parameters. 

\section*{Acknowledgments}
K.J. would like to thank Saurabh for very helpful discussions during the preparation of
this work.

\end{document}